\documentclass[a4paper,10pt]{article}
\usepackage{graphicx}
\usepackage{comment}
\usepackage{amssymb}
\usepackage{amsmath}
\usepackage{nicefrac}
\usepackage{graphicx}
\usepackage{dcolumn}
\usepackage{bm}
\usepackage{comment}
\usepackage{multirow}

\begin{document}

\huge

\begin{center}
Electron broadening operator including penetrating
collisions for hydrogen
\end{center}

\vspace{0.5cm}

\large

\begin{center}
Jean-Christophe Pain\footnote{jean-christophe.pain@cea.fr} and Franck Gilleron
\end{center}

\normalsize

\begin{center}
\it CEA, DAM, DIF, F-91297 Arpajon, France
\end{center}

\vspace{0.5cm}

\begin{abstract}
The expression of the electron broadening operator including the effect of penetrating collisions, \emph{i.e.} for which the incoming electron enters the extent of bound-electron wavefunctions, is rather complicated, even for hydrogen. It involves integrals of special functions, which evaluation deserves scrutiny. We present a simple approximate form of the electron collision operator for hydrogen including penetration effects, both in direct and interference terms. The new expression is accurate and easy to compute. In the Penetration Standard Theory, the collision operator is convergent whatever the value of the maximum impact parameter. However, when penetration theory is not valid anymore, it should be questioned. We discuss the problem of strong collisions when penetration effects are taken into account.
\end{abstract}

\newcommand{\bin}[2]{\left(\begin{array}{c}\!#1\!\\\!#2\!\end{array}\right)}
\newcommand{\ThreeJ}[6]{\left(\begin{array}{ccc}#1&#2&#3\\#4&#5&#6\end{array}\right)}
\newcommand{\SixJ}[6]{\left\{\begin{array}{ccc}#1&#2&#3\\#4&#5&#6\end{array}\right\}}
\newcommand{\M}{\mathcal{P}}
\newcommand{\F}{\mathcal{F}}
\newcommand{\Rnl}[2]{R_{#1}^{#2}}

\section{Introduction}\label{sec1}

Line-shape profiles are important ingredients of opacity and emissivity calculations, as they often serve as a diagnostics of laboratory or astrophysical plasmas. Indeed, the profiles contain information about local electric fields produced by electron and ion perturbers, leading to Stark splitting, and about density and temperature of the plasma. High mobility electrons perturb the emitter by collisions, possibly causing the interruption of the spontaneous emission and altering the emitter energy levels \cite{BANAZ07}. 

The problem of spectral line broadening due to emitter-perturber interactions has been largely studied. It started early with the works of Baranger \cite{BARANGER58b}, Kolb and Griem \cite{KOLB58} and Anderson's theory \cite{ANDERSON49}. During the last decades, the quantum statistical approach has been introduced to determine the shift and the width of spectral line shapes \cite{HITZSCHKE86,GUNTER95}. It is based on Green's function technique in which the line profiles are described by the two-particle polarization function related to the dipole-dipole correlation function. Besides the theoretical approaches, the computer simulation methods based on the molecular-dynamics approach, have been successfully applied to calculate the spectral line shapes \cite{IGLESIAS19}. In these computations, which are very efficient but expensive, the time evolution operator for the simple model of the plasma is obtained by solving numerically the time-dependent Schr\"odinger equation accounting for the many-body interactions between the emitter and the surrounding moving particles. Then, the spectral profile is obtained by averaging over a large number of plasma configurations \cite{STAMM79}. A few years ago, Bedida \emph{et al.} \cite{BEDIDA14} applied the path-integral formulation \cite{FEYNMAN65} to find the expression of the dipolar auto-correlation function in order to study the spectral line shapes in plasmas.

In the so-called ``standard line shape theory'' (ST: Standard Theory), electrons are modeled in a binary-collision theory, using classical path trajectory and often second-order perturbative treatment of electron broadening operator (Dyson series) \cite{SAHAL69a,SAHAL69b,DEUTSCH69a,DEUTSCH69b}. Ions are described in the quasi-static approximation. For both electrons and ions, the emitter-perturber interaction is often assumed to be dipolar only. It is permissible to separate the contributions of different frequency components into a fast and a slow component (as commonly done for electrons and ions) and convolve the resulting profile, as long as the fast component satisfy the impact approximation \cite{ALEXIOU13}. An issue in the standard electron treatment concerns the so-called strong collisions, \emph{i.e.} collisions associated to small impact parameters (or small electron velocities), for which perturbation theory is not valid and the dipole interaction is questionable due to penetration by the perturbing electrons into the atomic (bound-state) wave-function extent. A few years ago, it was suggested \cite{GRIEM97,ALEXIOU01,ALEXIOU05,ALEXIOU06,POQUERUSSE06,ALEXIOU17a,ALEXIOU17b,PAIN19} that penetration was likely to be more important than thought because the standard cutoff $n^2/Z$ ($n$ is the principal quantum number and $Z$ the atomic number) representing the wave-function extent in atomic units was too optimistic. When the spatial extent of the radiating states is comparable to the shielding length, or in other words when the important collisions with plasma electrons occur at distances within the extent of the wavefunctions of the levels involved in the line emission, collisions cannot be properly treated by the usual dipole, long-range approximation, which ``softens'' the interaction and reduces the widths \cite{GRIEM97b,GRIEM00}. A consequence of this softening of the interaction is that perturbation theory may remain valid, even for some collisions previously considered as strong \cite{ALEXIOU05}. In the Penetration Standard Theory, the collision operator is convergent whatever the value of the maximum impact parameter. However, the validity of penetration theory should be questioned as well. The problem of strong collisions when penetration effects are taken into account deserves scrutiny.

In the present work, we propose an approximate expression of the collision operator taking penetration effects into account. The study is restricted to the hydrogen atom, \emph{i.e.} straight-path trajectories. Even for hydrogen, the expression of the electron broadening operator including penetration effects is a difficult task. The formalism proposed by Alexiou and Poqu\'erusse \cite{ALEXIOU05} involves particular functions that the authors obtain from recurrence relations, initialized with Bessel and Bickley-Naylor functions. We found an exact expression of such functions, which enabled us to derive a simple approximate form of the electron collision operator for hydrogen including penetration effects, both in direct and interference terms. The new expression is accurate and easy to compute. In Sec. \ref{sec2}, the expression of the collision operator is recalled, and the factor representing the penetration effects is introduced. In Sec. \ref{sec3}, our new expression of the collision operator is presented and its applicability in the velocity - impact parameter domain is studied in Sec. \ref{sec4}. Section \ref{sec5} is the conclusion.
 
\section{The collision operator}\label{sec2}

\subsection{General form}\label{subsec21}

Throughout the paper, we set: $e=\hbar=m_e=1$ (atomic units) and $1/(4\pi\epsilon_0)=1$. In the standard theory, the matrix elements of the electron collision operator $\Phi$ read

\begin{eqnarray}
\langle\langle\alpha\beta|\Phi_{ab}|\alpha'\beta'\rangle\rangle=\sum_{\alpha''}\mathbf{r}_{\alpha\alpha''}.\mathbf{r}_{\alpha''\alpha'}~\phi_{\alpha\alpha'',\alpha''\alpha'}+\sum_{\beta''}\mathbf{r}_{\beta'\beta''}.\mathbf{r}_{\beta''\beta}~\phi_{\beta'\beta'',\beta''\beta}-\mathbf{r}_{\alpha\alpha'}.\mathbf{r}_{\beta'\beta}~\phi^{\mathrm{int}}_{\alpha\alpha',\beta'\beta},
\end{eqnarray}

\noindent where $\alpha$ and $\alpha'$ are upper level states, $\alpha''$ is a state perturbing the upper level states and $\mathbf{r}_{ij}$ are matrix elements of the position operator. $\Phi_{ab}$ is a tetradic (quadruply-indexed) operator, acting on initial subspace $a$ characterized by quantum numbers of states $\alpha$ and $\beta$ (\emph{i.e.} $\{n_{\alpha},\ell_{\alpha},m_{\alpha}\},\{n_{\beta},\ell_{\beta},m_{\beta}\}$) and on final subspace $b$ characterized by quantum numbers of states $\alpha'$ and $\beta'$ (\emph{i.e.} $\{n_{\alpha'},\ell_{\alpha'},m_{\alpha'}\},\{n_{\beta'},\ell_{\beta'},m_{\beta'}\}$). Spin quantum numbers are disregarded in the following. $\phi^{\mathrm{int}}$ is the interference term and $\phi$ (and $\phi^{\mathrm{int}}$) are velocity integrated complex functions of standard theory. More precisely, one has, choosing explicitly a straight line trajectory $\mathbf{R}(t)=\boldsymbol{\rho}+\mathbf{v}t$: 

\begin{eqnarray}
\phi_{\alpha\alpha'',\alpha''\alpha'}=\frac{\pi n_e}{3}\int vf(v)dv\int\rho \mathcal{I}_0\left(\rho,v;n_{\alpha},\ell_{\alpha},n_{\alpha''},\ell_{\alpha''}\right)\mathcal{I}_0\left(\rho,v;n_{\alpha''},\ell_{\alpha''},n_{\alpha'},\ell_{\alpha'}\right)d\rho,
\end{eqnarray}

\noindent where 

\begin{equation}
\mathcal{I}_0(\rho,v;n,\ell,n',\ell')=\rho\int_{-\infty}^{\infty}\frac{dt}{\left(\rho^2+v^2t^2\right)^{3/2}}.
\end{equation}

More than thirty years ago, a discussion took place about the physical meaning of the so-called interference term \cite{HEY75,VOSLAMBER76,GRIEM76,GIGOSOS07,GALTIER13}. That term does not account for any physical requirement: it results from a mathematical expansion in power series. It is a consequence of the fact that the states of the upper and lower groups feel the same perturbing field. This keeps coherence in the evolution of those states and reduces the broadening effect due to the collisions.

A collision with a plasma electron has a non-negligible probability amplitude to cause a transition $\alpha\rightarrow\alpha''$. $\beta$ and $\beta'$ are lower level states and $\beta''$ perturbs them. The no-quenching approximation consists in assuming that $\alpha,\alpha',\alpha''$ have the upper-level principal quantum number $n_{\alpha}=n_{\alpha'}=n_{\alpha''}$ and $\beta,\beta',\beta''$ have the lower-level principal quantum number $n_{\beta}=n_{\beta'}=n_{\beta''}$. In the present work, transitions due to collisions between the states of the upper group and the states of the lower group have not been taken into account. We use the notations

\begin{equation}
\sum_{i}\equiv\sum_{\ell_{i}=0}^{n_i-1}\sum_{m_{i}=-\ell_i}^{\ell_i}
\end{equation}

\noindent and

\begin{equation}\label{rij}
\mathbf{r}_{ij}=\langle n_i\ell_im_i|\vec{r}|n_j\ell_jm_j\rangle,
\end{equation}

\noindent where $\vec{r}=(x,y,z)$ is the position of the electron with respect to the center of the atom in cartesian coordinates. One has

\begin{equation}
\langle n\ell m|C_q^{(1)}|n\ell'm'\rangle=(-1)^{\ell-m}\ThreeJ{\ell}{1}{\ell'}{-m}{q}{m'}\langle\ell||C^{(1)}||\ell'\rangle \Rnl{n\ell}{n'\ell'}
\end{equation}

\noindent and

\begin{equation}
\langle\ell||C^{(1)}||\ell'\rangle=(-1)^{\ell}\sqrt{(2\ell+1)(2\ell'+1)}\ThreeJ{\ell}{1}{\ell'}{0}{0}{0}.
\end{equation}

\noindent One has also

\begin{equation}
\left\{
\begin{array}{l}
x=\frac{1}{\sqrt{2}}\left(C_{-1}^{(1)}-C_{1}^{(1)}\right)r\\
y=\frac{i}{\sqrt{2}}\left(C_{-1}^{(1)}+C_{1}^{(1)}\right)r\\
z=C_0^{(1)}r.
\end{array}
\right.
\end{equation}

\noindent The general formula for dipole ($|\ell-\ell'|=1$) radial integrals $\Rnl{n\ell}{n'\ell'}$ of one-electron systems has been obtained by Gordon in terms of hypergeometric functions \cite{GORDON29,BETHE57,TARASOV03}. For $n'\ne n$, one has\footnote{the formula has been symmetrized to account for $\ell'=\ell\pm 1$.}:

\begin{align}
\Rnl{n\ell}{n'\ell'}=\Rnl{n'\ell'}{n\ell}=&\frac{(-1)^{n'-\ell_>}}{4Z(\ell+\ell')!}~\sqrt{\frac{(n+\ell)!(n'+\ell')!}{(n-\ell-1)!(n'-\ell'-1)!}}~X^{1+\ell_>}~Y^{n+n'}\nonumber\\
&\left[
\phantom{}_2F_1(-n+\ell+1,-n'+\ell'+1;2\ell_>;-X)
-\phantom{}_2F_1(-n+\ell',-n'+\ell;2\ell_>;-X)Y^2
\right],
\end{align}

\noindent where $\ell_>=\max(\ell,\ell')$, $X=4nn'/(n-n')^2$ and $Y=(n-n')/(n+n')$. The Gauss hypergeometric function is:

\begin{align}
\phantom{}_2F_1(a,b;c;z)=\sum_{k=0}^ \infty \bin{a+k-1}{k}\bin{b+k-1}{k} \bin{c+k-1}{k}^{-1}z^k.
\end{align}

\noindent In the case $n'=n$, the formula is simpler:

\begin{align}
\Rnl{n\ell}{n\ell'}=-\frac{3}{2Z}n\sqrt{n^2-\ell_>^2}.
\end{align}

\subsection{Collision integral}\label{subsec22}
 
The impact approximation is valid when \cite{SAHAL69a,SAHAL69b}:

\begin{itemize}

\item The duration of a collision is small compared to the mean time between collisions. In that case, radiation can be neglected during the collision, which can be considered as instantaneous. 

\item The duration of a collision is much smaller than the inverse HWHM (half width at half maximum) of the profile $\Delta\omega$.

\item The collisions are complete, which means that they can be considered as instantaneous in comparison with $\Delta\omega^{-1}$. Therefore, the radiation process of the emitter can be decoupled from the interaction process with perturbers.

\end{itemize}

In fact, an impact theory with a complete-collision assumption can be used only for values of $\Delta \omega$ smaller than the electron plasma frequency. It was shown that the complete-collision assumption may be corrected in the line wings by means of the Lewis cutoff \cite{LEWIS60}. The so-called relaxation theory \cite{SMITH68} does not make such an assumption, and is in good agreement with the impact theory corrected by the Lewis cutoff.

In the classical picture, the electrons are assumed to follow straight paths for the hydrogen \cite{GRIEM59} and neutral helium lines \cite{GRIEM62}, while the hyperbolic trajectories must be used when the lines are emitted by ions \cite{GRIEM61,KEPPLE72}. The classical path assumption for hydrogen yields results which are identical to the quantum-mechanical ones \cite{BARANGER58a}. The theory of hydrogen line broadening by electrons must take into account the non-adiabatic nature of the perturbation. Collisional transitions between the states of the same shell play the main role in the broadening. The broadening of the spectral line
 due to the collision with the atomic electron involved in the transition between $a$ (states $\alpha$, $\alpha'$, etc.) and $b$ (states $\beta$, $\beta'$, etc.) can be expressed in terms of the $S$-matrix elements. To simplify the scattering by $N$ electrons, one assumes that an electron comes very close to the atom, which creates a huge electric field, and the electric field of all the other electrons can be neglected compared to that of the close one. Hence, the total electric field can be replaced by the field of a single electron and the result then is multiplied by the number of electrons \cite{GRIEM62}. According to Griem, this should be done only in one of the two fields entering the second-order term. In the other one, the total field must be used, approximated by a screened effective field. The matrix element of the tetradic collision operator reads

\begin{eqnarray}
\langle\langle\alpha\beta|\Phi_{ab}|\alpha'\beta'\rangle\rangle=n_e\int vf(v)dv\int 2\pi\rho d\rho\langle\langle\alpha\beta|\left\{1-S_aS_b^{\dag}\right\}|\alpha'\beta'\rangle\rangle,
\end{eqnarray}

\noindent where $n_e$ is the electron density, $f(v)$ represents the velocity ($v$) distribution of the perturber and $\rho$ the impact parameter. The braces $\{\}$ denote the angular averaging, \emph{i.e.} the averaging over directions of vectors $\rho$ and $v$, and $S_a$ and $S_b$ are the scattering $S$ matrices for collisions with the atom or ion being in $a$ or $b$ state, respectively. We have used the notation

\begin{equation}
\langle\langle\alpha\beta|S_aS_b^{\dag}|\alpha'\beta'\rangle\rangle=\langle\alpha|S_a|\alpha'\rangle\langle\beta|S_b^{\dag}|\beta'\rangle.
\end{equation}

The derivation of the collision operator in the interaction picture in the general case is is briefly recalled in Appendix \ref{appendixa}, as well as the particular case of hydrogen.

\subsection{Collision integral $\mathcal{I}$ and factor $C_1$ accounting for penetration}\label{subsec23}

In the penetrating standard theory, we have, choosing explicitly a straight line trajectory $\mathbf{R}(t)=\boldsymbol{\rho}+\mathbf{v}t$: 

\begin{eqnarray}
\phi_{\alpha\alpha'',\alpha''\alpha'}=\frac{\pi n_e}{3}\int vf(v)dv\int\rho \mathcal{I}\left(\rho,v;n_{\alpha},\ell_{\alpha},n_{\alpha''},\ell_{\alpha''}\right)\mathcal{I}\left(\rho,v;n_{\alpha''},\ell_{\alpha''},n_{\alpha'},\ell_{\alpha'}\right)d\rho,
\end{eqnarray}

\noindent where $n_i$ and $\ell_i$ are respectively the principal and orbital quantum numbers of state $i$. The integral

\begin{equation}
\mathcal{I}(\rho,v;n,\ell,n',\ell')=\rho\int_{-\infty}^{\infty}\frac{C_1\left(n,\ell,n',\ell';\sqrt{\rho^2+v^2t^2}\right)}{\left(\rho^2+v^2t^2\right)^{3/2}}dt
\end{equation}

\noindent essentially includes the atomic-collision physics and $C_1$ is a factor accounting exactly for penetration in the dipolar approximation. It is a particular case of $C_{\lambda}$ ($\lambda$ is actually the multipolarity). The standard behavior is recovered if $C_{\lambda}=1$ (no penetration) and in that case $\mathcal{I}=2/(\rho v)$. The origin is taken at the location of the emitter. If $\mathbf{r}$ is the position of the bound electron and $\mathbf{R}$ the position of the incoming electron, the Coulomb interaction energy is:

\begin{equation}
V=\frac{1}{|\mathbf{r}-\mathbf{R}|}-\frac{1}{|\mathbf{R}|},
\end{equation}

\noindent and

\begin{equation}
\frac{1}{|\mathbf{r}-\mathbf{R}|}=\sum_{k=0}^{\infty}\frac{r_<^{k}}{r_>^{k+1}}P_k(\cos\theta),
\end{equation}

\noindent where $\mathbf{r}.\mathbf{R}=rR\cos\theta$, $r_<=\min(r,R)$, $r_>=\max(r,R)$ and $P_k$ is Legendre polynomial \cite{NGUYENHOE64,DEUTSCH73,GOMEZ16}. Neglecting penetration ($R\gg r$):

\begin{equation}
V=\underbrace{\frac{1}{R}}_{\mathrm{monopole}}-\frac{1}{R}+\underbrace{\left(\frac{\mathbf{r}.\mathbf{R}}{R^3}\right)}_{\mathrm{dipole}}+\underbrace{\left(3\frac{(\mathbf{r}.\mathbf{R})^2}{2R^5}-\frac{r^2}{2R^3}\right)}_{\mathrm{quadrupole}}+\cdots.
\end{equation}

\noindent If penetration is taken into account (dipolar term only):

\begin{equation}
V=\frac{r_<}{r_>^2}P_1(\cos\theta)=\frac{r_<}{r_>^2}\frac{\mathbf{r}.\mathbf{R}}{rR},
\end{equation}

\noindent \emph{i.e.}

\begin{equation}
V=\frac{r_<}{r_>^2}\frac{R^2}{r}\mathbf{r}.\frac{\mathbf{R}}{R^3}.
\end{equation}

\noindent Therefore, the radial dipolar integrals are modified by a multiplicative factor $C_1(R)$ which reads

\begin{equation}
C_1(R)=\frac{\int_0^{\infty}P_{n\ell}(r)P_{n'\ell'}(r)\frac{r_<}{r_>^2}R^2dr}{\int_0^{\infty}P_{n\ell}(r)P_{n'\ell'}(r)rdr},
\end{equation}

\noindent \emph{i.e.}

\begin{equation}
C_1(R)=\frac{\int_0^{R}P_{n\ell}(r)P_{n'\ell'}(r)rdr}{\int_0^{\infty}P_{n\ell}(r)P_{n'\ell'}(r)rdr}+R^3\frac{\int_R^{\infty}P_{n\ell}(r)P_{n'\ell'}(r)\frac{1}{r^2}dr}{\int_0^{\infty}P_{n\ell}(r)P_{n'\ell'}(r)rdr},
\end{equation}

\noindent where $P_{n\ell}(r)$ is the radial part of the wave-function multiplied by $r$. In the more general case, taking into account all the multipolarities, we have

\begin{equation}
V=\sum_{\lambda}\left(\frac{r_<^{\lambda}}{r_>^{\lambda+1}}\frac{R^{\lambda+1}}{r^{\lambda}}\right)\left(\frac{r^{\lambda}}{R^{\lambda+1}}P_{\lambda}\left(\frac{\mathbf{r}.\mathbf{R}}{rR}\right)\right)
\end{equation}

\noindent and therefore we get, for the coefficient $C_{\lambda}$, taking into account the correction to the multipolar integral of order $\lambda$:

\begin{equation}
C_{\lambda}(R)=\frac{\int_0^{\infty}P_{n\ell}(r)P_{n'\ell'}(r)\frac{r_<^{\lambda}}{r_>^{\lambda+1}}R^{\lambda+1}dr}{\int_0^{\infty}P_{n\ell}(r)P_{n'\ell'}(r)r^{\lambda}dr},
\end{equation}

\noindent \emph{i.e.}

\begin{equation}
C_{\lambda}(R)=\frac{\int_0^{R}P_{n\ell}(r)P_{n'\ell'}(r)r^{\lambda}dr}{\int_0^{\infty}P_{n\ell}(r)P_{n'\ell'}(r)r^{\lambda}dr}+R^{2\lambda+1}\frac{\int_R^{\infty}P_{n\ell}(r)P_{n'\ell'}(r)\frac{1}{r^{\lambda+1}}dr}{\int_0^{\infty}P_{n\ell}(r)P_{n'\ell'}(r)r^{\lambda}dr}.
\end{equation}

\subsection{Consequences of penetration}\label{subsec24}

Penetration usually ``softens'' the interaction in the sense that it tends to reduce the broadening, at least for isolated lines \cite{ALEXIOU01}. However, it was shown that in some cases, especially for strong coupling conditions, penetration can enhance the broadening \cite{ALEXIOU17a}, when small impact parameters are involved and when the shielding length becomes of the same order as the wave-function extent (\emph{e.g.} in the case of line merging \cite{INGLIS39}).

\section{Approximate form of the collision operator}\label{sec3}

\subsection{Collision integral and function $\Delta(b)$}\label{subsec31}

The collision integral $\mathcal{I}$ can be put in the form

\begin{equation}
\mathcal{I}=\frac{2}{\rho v}\left[1-\Delta(b)\right]
\end{equation}

\noindent with $b=2\rho/n$ and, in the dipolar case: 

\begin{equation}\label{truedelta}
\Delta(b)=\sum_{i=2}^{2n+\lambda}s_ib^iF_{i-2}(b)+bK_1(b).
\end{equation}

\noindent The coefficients $s_i$, which are rapidly decreasing functions of $i$, can be computed exactly and are provided in the Appendix of Ref. \cite{ALEXIOU05}. We recently published the explicit forms \cite{PAIN19}:

\begin{eqnarray}\label{feven}
F_{2p}(b)&=&\frac{1}{2^{2p}}\left\{2\sum_{k=0}^{p-1}\bin{2p}{k}K_{2p-2k}\left(b\right)+\bin{2p}{p}K_0(b)\right\}
\end{eqnarray}

\noindent and

\begin{equation}\label{fodd}
F_{2p+1}(b)=\frac{1}{2^{2p}}\sum_{k=0}^{p}\bin{2p+1}{k}K_{2p-2k+1}\left(b\right),
\end{equation}

\noindent which do not require to resort to recurrence relations. The function $\phi$ then reads

\begin{eqnarray}
\phi_{\alpha\alpha'',\alpha''\alpha'}=\frac{4\pi n_e}{3}\sqrt{\frac{2}{\pi k_BT}}\int_0^{b_{\mathrm{max}}}\frac{db}{b}\left[1-\Delta\left(b;n_{\alpha},\ell_{\alpha},\ell_{\alpha''}\right)\right]\left[1-\Delta\left(b;n_{\alpha},\ell_{\alpha''},\ell_{\alpha'}\right)\right]
\end{eqnarray}

\noindent and

\begin{eqnarray}
\phi^{\mathrm{int}}_{\alpha\alpha',\beta'\beta}=\frac{8\pi n_e}{3}\sqrt{\frac{2}{\pi k_BT}}\int_0^{b_{\mathrm{max}}}\frac{db}{b}\left[1-\Delta\left(b;n_{\alpha},\ell_{\alpha},\ell_{\alpha'}\right)\right]\left[1-\Delta\left(\frac{n_{\alpha}b}{n_{\beta}};n_{\beta},\ell_{\beta'},\ell_{\beta}\right)\right],
\end{eqnarray}

\noindent where $b_{\mathrm{max}}=2\rho_{\mathrm{max}}/n_{\alpha}$ is a cutoff introduced to avoid the logarithmic divergence of the integral at large impact parameters ($\Delta(b)\rightarrow 0$ when $b\rightarrow\infty$). As for the standard theory, the maximum impact parameter $\rho_{\mathrm{max}}$ is usually chosen to be of the order of the Debye length 

\begin{equation}
\lambda_D=\sqrt{\frac{k_BT}{4\pi n_e}}
\end{equation}

\noindent or 1.1$\lambda_D$ (respectively 0.68$\lambda_D$) to account for the single \cite{GRIEM62} (respectively double \cite{CHAPPELL69}) shielded fields in the $S$-matrix. Table \ref{deb} gives the value of $\lambda_D$ for different plasma conditions as well as the approximate value of principal quantum number $n$ such that $n^2a_0\approx\lambda_D$. We can see that low-$n$ shells can be concerned with penetration theory, especially in the interior of the Sun, but one must keep in mind that in such cases the Debye length is not a good estimate of the screening length, and it may be more relevant to choose the Thomas-Fermi length. Moreover, due to pressure ionization, the maximum value of $n$ is determined by the density (in the three first cases: gas discharge, tokamak and ionosphere).

\begin{table}[h]
\centering
\begin{tabular}{|c|c|c|c|c|}\hline
Plasma & Electron & Electron & Debye & Approximate value   \\
 & density $n_e$ (m$^{-3}$) & temperature & length $\lambda_D$ (m) & of $n$ such that $n^2a_0\approx\lambda_D$ \\
 & & & & (case of hydrogen)  \\\hline
Gas discharge & 10$^{16}$ & 10$^{4}$ & 7 10$^{-5}$ & 1000 \\ 
Tokamak & 10$^{20}$ & 10$^{8}$ & 7 10$^{-5}$ & 1000 \\
Ionosphere & 10$^{12}$ & 10$^{3}$ & 2 10$^{-3}$ & 6000 \\ 
Solar center & 10$^{32}$ & 10$^{7}$ & 2 10$^{-11}$ & 1 \\
Half-radius & 4 10$^{29}$ & 3 10$^{6}$ & 2 10$^{-10}$ & 2 \\
of the Sun & & & & \\
Base of the convective & 10$^{28}$ & 2 10$^{6}$ & 10$^{-9}$ & 4 \\
zone of the Sun & & & & \\\hline
\end{tabular}
\caption{Debye length and approximate value of $n$ such that $n^2a_0=\lambda_D$ (case of hydrogen).}
\label{deb}
\end{table}

\subsection{Analytical representation of the collision operator}\label{subsec32}

It is possible to obtain simple approximate formula for $\Delta(b)$ which integral gives the collision operator. Noticing that the quantity $\Delta(b)$ has a half-bell shape with $\Delta(0)=1$ and $\Delta(\infty)=0$, we tried to find an approximation with the function

\begin{equation}\label{appd}
\Delta_{\mathrm{app}}\left(b;n,\ell,\ell'\right)=\exp\left[-\frac{b^2}{2\chi_{n,\ell,\ell'}^2}\right],
\end{equation}

\noindent with

\begin{equation}\label{chinl}
\chi_{n,\ell,\ell'}=\frac{\sqrt{2\pi}}{8}~\frac{\left[5n^2-\ell_<\left(\ell_<+2\right)\right]}{n},
\end{equation}

\noindent where $\ell_<=\min(\ell,\ell')$ and $\ell_>=\max(\ell,\ell')$. The approximant of $\Delta(b)$ provides an approximate expression for the collision operator itself. Let us consider for instance the term

\begin{eqnarray}
\phi_{\alpha\alpha'',\alpha''\alpha}=\frac{4\pi}{3}n_e\sqrt{\frac{2}{\pi k_BT}}\int_0^{b_{\mathrm{max}}}\frac{\left[1-\Delta\left(b;n_{\alpha},\ell_{\alpha},\ell_{\alpha''}\right)\right]^2}{b}db,
\end{eqnarray}

\noindent In our previous work \cite{PAIN19}, we only provided the expression of the following part of the collision operator\footnote{The is a typo in the caption of Figure 6 in Ref. \cite{PAIN19}: the end of the last sentence should read ``$n_{\alpha}$=3, $\ell_{\alpha}$=2 and $\ell_{\alpha''}$=1.'' (but the legend is correct).}:

\begin{equation}\label{diag}
\phi_{\alpha\alpha'',\alpha''\alpha}=\frac{4\pi}{3}n_e\sqrt{\frac{2}{\pi k_BT}}~\tilde{\phi}_{\alpha\alpha''},
\end{equation}

\noindent with

\begin{equation}\label{diagapp}
\tilde{\phi}_{\alpha\alpha''}=G\left(\frac{16\lambda_D}{\sqrt{2\pi}\left[5n^2-\ell_<(\ell_<+2)\right]}\right),
\end{equation}

\noindent and

\begin{equation}
G(x)=\frac{\gamma_E}{2}-\frac{1}{2}E_1\left(x^2\right)+E_1\left(\frac{x^2}{2}\right)+\ln\left(\frac{x}{2}\right),
\end{equation}

\noindent where $\gamma_E$ is the Euler constant and $E_1$ the exponential integral:

\begin{equation}
E_1(x)=\int_x^{\infty}t^{-1}\exp(-t)dt
\end{equation}

\begin{figure}
\begin{center}
\vspace{1cm}
\includegraphics[width=10cm]{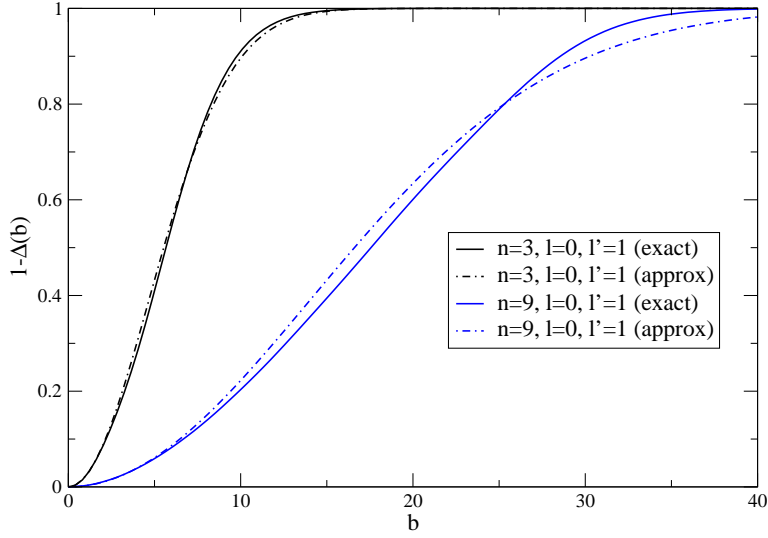}
\end{center}
\caption{Comparison between the exact computation of $\left[1-\Delta(b)\right]$ (Eq. (\ref{truedelta}), (\ref{fodd}) and (\ref{feven})) and the approximate expression (\ref{appd}) for $\ell=0$, $\ell'=1$ and two values of principal quantum number $n$: 3 and 9.}\label{approx}
\end{figure}

\begin{figure}
\begin{center}
\vspace{1cm}
\includegraphics[width=10cm]{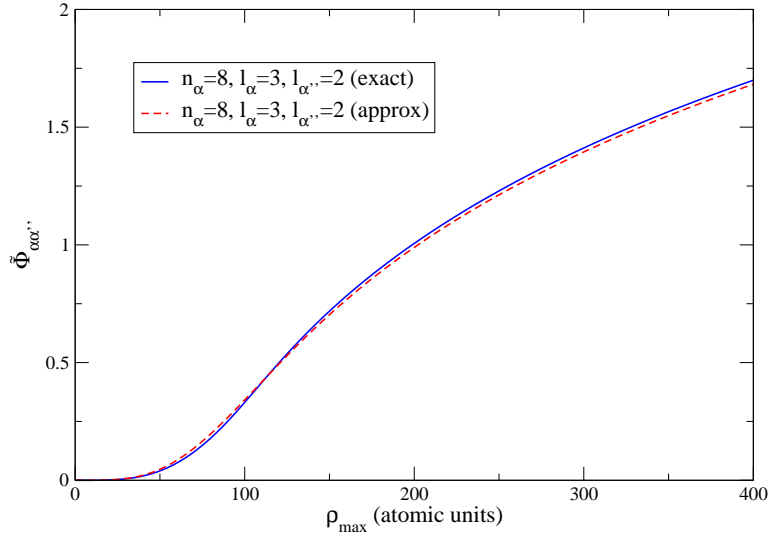}
\end{center}
\caption{Comparison between the exact computation of the quantity $\tilde{\Phi}_{\alpha\alpha''}$ entering the diagonal part of the collision operator (see Eq. (\ref{diag})) and the approximate expression (\ref{diagapp}) for $n_{\alpha}=8$, $\ell_{\alpha}=3$ and $\ell_{\alpha''}=2$.}\label{op_coll_ex2}
\end{figure}

\noindent In fact, the most general case is the integral

\begin{equation}
g(x_1,x_2)=\int_0^{b_{\mathrm{max}}}\frac{\left(1-\exp\left[-\frac{b^2}{2b_{\mathrm{max}}^2}x_1^2\right]\right)\left(1-\exp\left[-\frac{b^2}{2b_{\mathrm{max}}^2}x_2^2\right]\right)}{b}db,
\end{equation}

\noindent which is equal to

\begin{equation}
g(x_1,x_2)=\frac{\gamma_E}{2}+\frac{1}{2}E_1\left(\frac{x_1^2}{2}\right)+\frac{1}{2}E_1\left(\frac{\lambda^2x_2^2}{2}\right)-\frac{1}{2}E_1\left(\frac{x_1^2+\lambda^2x_2^2}{2}\right)+\ln\left(\frac{x_1^2x_2^2\lambda^2}{2\left(x_1^2+x_2^2\lambda^2\right)}\right),
\end{equation}

\noindent and the expression of the collision operator is

\begin{eqnarray}\label{newfor}
\langle\langle\alpha\beta|\Phi|\alpha'\beta'\rangle\rangle&=&\frac{4\pi}{3}n_e\sqrt{\frac{2}{\pi k_BT}}\sum_{\alpha''}\mathbf{r}_{\alpha\alpha''}.\mathbf{r}_{\alpha''\alpha'}~g(x_{\alpha\alpha''},x_{\alpha''\alpha'},1)+\sum_{\beta''}\mathbf{r}_{\beta'\beta''}.\mathbf{r}_{\beta''\beta}~g(x_{\beta'\beta''},x_{\beta''\beta},1)\nonumber\\
& &-\mathbf{r}_{\alpha\alpha'}.\mathbf{r}_{\beta'\beta}~g(x_{\alpha\alpha'},x_{\beta'\beta},\frac{n_{\alpha}}{n_{\beta}}),
\end{eqnarray}

\noindent where

\begin{equation}
x_{ij}=\frac{b_{\mathrm{max}}}{\chi_{n,\ell_i,\ell_j}}
\end{equation}

\noindent with $n=n_{\alpha}=n_{\alpha'}=n_{\alpha''}$, $n_{\beta}=n_{\beta'}=n_{\beta''}$ and $\{i,j\}\in\left(\alpha, \alpha',\alpha'',\beta,\beta',\beta''\right)$. Figure \ref{approx} shows a comparison between the exact computation of $\left[1-\Delta(b)\right]$ (Eq. (\ref{truedelta}), (\ref{fodd}) and (\ref{feven})) and the approximate expression (\ref{appd}) for $\ell=0$, $\ell'=1$ and two values of principal quantum number $n$: 3 and 9. But can see that the agreement is good. In the same way, figure \ref{op_coll_ex2} displays a comparison between the exact computation of the quantity $\tilde{\Phi}_{\alpha\alpha''}$ entering the diagonal part of the collision operator (see Eq. (\ref{diag})) and the approximate expression (\ref{diagapp}) for $n_{\alpha}=8$, $\ell_{\alpha}=3$ and $\ell_{\alpha''}=2$. We can see that the two calculations are almost superimposed.

\section{Strong collisions}\label{sec4}

In many approaches to shift and broadening of spectral lines, a low-order perturbation treatment has been used for the interaction between the radiator and the perturbing electrons. However, such a treatment is allowed for weak collisions only. Dealing with strong collisions \cite{BACON68,GRIEM79,ALEXIOU09}, a low-order perturbative treatment even for the electron-atom interaction leads to an overestimation of strong-collision contributions \cite{GUNTER93}. Within a semi-classical treatment of the electron-radiator collisions, a low-order perturbative expansion produces even divergent integrals for shift and width. Although it is possible to overcome these divergences within a full quantum theory, contributions of strong electron-atom collisions will be overestimated further on. Therefore, in earlier papers (see for instance Refs. \cite{GUNTER90,GUNTER91}), a simple cutoff procedure, as proposed by Griem \cite{GRIEM74,GRIEM83}, has been applied for strong collisions. However, such a procedure is not well founded from the theoretical point of view. Furthermore, the intrinsically non-unique choice of a cutoff parameter remains unsatisfactory. Whereas for linewidth calculations such a cutoff procedure has been proven to be successful, for the line shift a cutoff procedure is problematic \cite{IGLESIAS85}. Further, it remains an open question whether there are strong-collision contributions to the line shift at all. In Ref. \cite{GRIEM88} these contributions have been estimated to be about 20 \% of the weak-collision contributions. However, such an estimation could not be established yet. Of course, within the unified theories \cite{GRIEM88,VOSLAMBER69,VIDAL70} strong-collision contributions which do not overlap in time have been included. Unfortunately, due to the used no-quenching approximation, no line shifts could be calculated within this theory. Another way to deal with strong collisions is to make use of the well-known relation between shift and width of the line and the scattering phase shifts given by Baranger \cite{BARANGER58a,BARANGER58b,BARANGER59}. Thus the problem is transformed into the calculation of phase shifts for the electron scattering at excited atomic states. In properly dealing with this problem, usually many atomic states must be included into the following close-coupling equations. That is why it is difficult to carry out such phase-shift calculations, especially for highly excited atomic states, although interesting work has been done for determining shift and width for the first hydrogen lines using asymptotic $S$-matrix elements \cite{UNNIKRISHNAN91}. As already shown in previous papers, a Green's-function approach is well suited to deal with spectral line shapes. Using the advantages of the diagram technique, one can find easily a complete set of contributing terms within a definite frame of approximations. G\"unter introduced a two-particle Green's function approach to get tractable expressions for shift and width of spectral lines including strong-collision contributions \cite{GUNTER93}. In the latter paper strong-collision contributions to the line shift have also been investigated. Thus the often used cutoff procedure for strong collision contributions introduced by Griem \cite{GRIEM64,GRIEM74,GRIEM83} could be replaced by an approach treating strong-collision contributions in a consequent manner. In order to test the developed theory, as an example, the shape of the hydrogen Lyman-$\alpha$ line has been calculated. The resulting line profile agrees excellently with the unified theory results \cite{GREENE79}. The calculated shift of the Lyman$_{\alpha}$ line is somewhat, smaller than it has been given by Griem \cite{GRIEM88}.

Unfortunately, benchmark experimental data are scarce for high-density plasmas, where both strong collisions and penetration effects are important (see for instance \cite{BODDEKER93,BUSCHER02}).   

\subsection{Case of Standard Theory}\label{subsec41}

We have seen that the collision operator can be put in the form

\begin{equation}\label{phidiv}
\Phi_{ab}=n_e\int vf(v)dv\int 2\pi\rho d\rho\left\{1-S_aS_b^{\dag}\right\},
\end{equation}

\noindent where $S_a$ and $S_b$ are the scattering matrices in the states $a$ and $b$. This results from the impact theory. Griem suggested to integrate first on the impact parameters and then on the velocities \cite{GRIEM74}. Considering the diagonal part of $\Phi_{ab}$, one has

\begin{equation}
\langle\langle\alpha\beta|\left\{1-S_aS_b^{\dag}\right\}(\rho,v)|\alpha\beta\rangle\rangle\approx\frac{2}{3\left(\rho v\right)^2}\left(\langle r_{\alpha}\rangle-\langle r_{\beta}\rangle\right)^2,
\end{equation}

\noindent where $\langle r_i\rangle=\langle i|r|i\rangle$. The integral over impact parameter in Eq. (\ref{phidiv}) thus yields a logarithmic divergence as $\rho$ tends to infinity. Since the electron does not feel the potential of the emitter beyond a certain distance, we introduce a cut-off $\rho_{\mathrm{max}}$ and the contribution of screened collisions to $\Phi_{ab}$ is assumed to be zero.

The integral over impact parameter in Eq. (\ref{phidiv}) also diverges as $\rho$ tends to zero. This is due to the fact that this expression of $\Phi_{ab}$ stems from a second-order perturbative treatment, and is valid only if unitarity is ensured, \emph{i.e.} $\langle\langle\alpha\beta|\left\{1-S_aS_b^{\dag}\right\}(\rho,v)|\alpha\beta\rangle\rangle\leq C$. $C$ is sometimes called a ``strong collision'' constant. The quantity $\left\{1-S_aS_b\right\}(\rho,v)$ is used to subdivide the collisions into weak and strong ones. The frontier is defined by the fact that the weak collisions correspond to

\begin{equation}
\frac{2}{3\rho^2v^2}\left(\langle r_{\alpha}\rangle-\langle r_{\beta}\rangle\right)^2\leq C,
\end{equation}

\noindent which means that

\begin{equation}
\rho\geq\rho_{\mathrm{w}}(v),
\end{equation}

\noindent where the Weisskopf radius $\rho_{\mathrm{w}}$ is 

\begin{equation}\label{weiss}
\rho_{\mathrm{w}}^2(v)\approx\frac{1}{C}\frac{2}{3v^2}\left(\langle r_{\alpha}\rangle-\langle r_{\beta}\rangle\right)^2.
\end{equation}

\noindent In terms of velocity, one has

\begin{equation}\label{vst}
v\leq v_{\mathrm{ST}}(\rho)=\sqrt{\frac{2}{3C}}\frac{1}{\rho}\left(\langle r_{\alpha}\rangle-\langle r_{\beta}\rangle\right)^2
\end{equation}

\noindent and the weak-collision term can therefore be treated as

\begin{equation}
\Phi_{ab}^{\mathrm{weak}}=n_e\int_{v_{\mathrm{ST}}(\rho_{\mathrm{max}})}^{\infty}vf(v)dv\int_{\rho_{\mathrm{w}}(v)}^{\rho_{\mathrm{max}}} 2\pi\rho d\rho\left\{1-S_aS_b^{\dag}\right\}.
\end{equation}

\begin{figure}
\begin{center}
\includegraphics[width=14.cm]{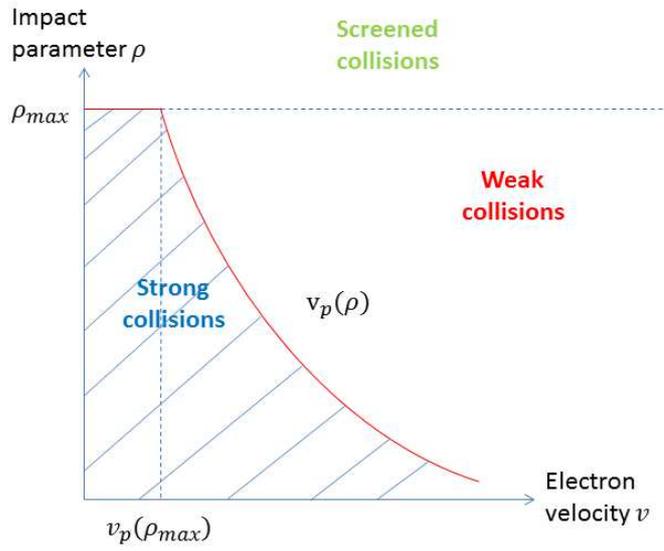}
\end{center}
\caption{Simplified schematic representation of three different collisional regimes: screened, weak and strong in the ($\rho$,$v$) plane, $\rho$ being the impact parameter and $v$ the electron velocity. $v_{\mathrm{p}}(\rho)$ represents the frontier between strong and weak collisions. It corresponds to $\left\{1-S_aS_b^{\dag}\right\}=1$.}\label{fig12}
\end{figure}

\noindent As indicated by Eq. (\ref{weiss}), the quantity $\rho_{\mathrm{w}}(v)$ depends on the choice of $C$. Kepple and Griem chose $C$=1 \cite{KEPPLE68}. Later in his book, Griem took $C$=3/2. Oks suggests $C\leq 2$ \cite{OKS15}. In the following, we set $C$=1, but the results can be generalized to other values of $C$. It is relevant to define three regimes (see Fig. \ref{fig12}):

\begin{itemize}

\item Weak collisions: $\int_0^{\rho_{\mathrm{max}}}d\rho\int_{v_{\mathrm{ST}}(\rho)}^{\infty}dv$\;\;\;\; or\;\;\;\;$\int_{v_{\mathrm{ST}}(\rho_{\mathrm{max}})}^{\infty}dv\int_{\rho_{\mathrm{w}}(v)}^{\rho_{\mathrm{max}}}d\rho$. Unitarity ensured, $\left\{1-S_aS_b^{\dag}\right\}$ in the integrand.

\vspace{1cm}

\item Strong collisions: $\int_0^{\rho_{\mathrm{max}}}d\rho\int_0^{v_{\mathrm{ST}}(\rho)}dv$\;\;\;\; or\;\;\;\;$\int_{v_{\mathrm{ST}}(\rho_{\mathrm{max}})}^{\infty}dv\int_0^{\rho_{\mathrm{w}}(v)}d\rho + \int_0^{v_{\mathrm{ST}}(\rho_{\mathrm{max}})}dv\int_0^{\rho_{\mathrm{max}}}d\rho$. Unitarity violated, $\left\{1-S_aS_b^{\dag}\right\}=C$ in the integrand (Lorentz-Weisskopf approach).

\vspace{1cm}

\item Screened collisions: $\int_{\rho_{\mathrm{max}}}^{\infty}d\rho\int_0^{\infty}dv$\;\;\;\; or\;\;\;\;$\int_0^{\infty}dv\int_{\rho_{\mathrm{max}}}^{\infty}d\rho$. No contribution to $\Phi_{ab}$.

\end{itemize} 

The diagonal element of the strong collision part is

\begin{equation}
\Phi_{\mathrm{strong,ST}}=n_e\left\{\int_{v_{\mathrm{ST}}\left(\rho_{\mathrm{max}}\right)}^{\infty}\int_0^{\rho_{\mathrm{w}}(v)} + \int_0^{v_{\mathrm{ST}}\left(\rho_{\mathrm{max}}\right)}\int_0^{\rho_{\mathrm{max}}}\right\}vf(v)dv~2\pi\rho d\rho
\end{equation}

\noindent and the minimum velocity is given by

\begin{equation}
\rho_{\mathrm{w}}(v)\leq\rho_{\mathrm{max}}\rightarrow v\ge v_{\mathrm{ST}}\left(\rho_{\mathrm{max}}\right).
\end{equation}

\noindent In the following, we replace $\langle r_{\alpha}\rangle$ by $n_{\alpha}^2$ but in section \ref{subsec42}, we use the exact non-relativistic expression for average quantities, depending on $n_{\alpha}$ and $\ell_{\alpha}$. Assuming the Maxwell distribution for the incoming free electron

\begin{equation}
f(v)=4\pi v^2\left(\frac{1}{2\pi k_BT}\right)^{3/2}\exp\left[-\frac{v^2}{2k_BT}\right],
\end{equation}

\noindent we get

\begin{equation}
\Phi_{\mathrm{strong,ST}}=2n_e\sqrt{2\pi k_BT}~\rho_{\mathrm{max}}^2\left\{1-\exp\left[-\frac{v_{\mathrm{ST}}^2\left(\rho_{\mathrm{max}}\right)}{2k_BT}\right]\right\}.
\end{equation}

\noindent Noting that

\begin{equation}
\langle\langle\alpha\beta|\left\{1-S_aS_b^{\dag}\right\}(\rho,v)|\alpha\beta\rangle\rangle=\frac{v_{\mathrm{ST}}^2(\rho)}{v^2},
\end{equation}

\noindent the weak-collision diagonal part is

\begin{equation}
\Phi_{\mathrm{weak,ST}}=n_e\int_{v_{\mathrm{ST}}\left(\rho_{\mathrm{max}}\right)}^{\infty}\int_{\rho_{\mathrm{w}}(v)}^{\rho_{\mathrm{max}}}vf(v)dv\frac{v_{\mathrm{ST}}^2(\rho)}{v^2}d\rho
\end{equation}

\noindent \emph{i.e.}

\begin{equation}
\Phi_{\mathrm{weak,ST}}=\frac{2n_e}{3}\sqrt{\frac{2\pi }{k_BT}}~E_1\left[\frac{v_{\mathrm{ST}}^2\left(\rho_{\mathrm{max}}\right)}{2k_BT}\right]\left(n_{\alpha}^2-n_{\beta}^2\right)^2
\end{equation}

\noindent and therefore the total diagonal matrix element of the collision operator reads

\begin{eqnarray}\label{phitotfirst}
\Phi_{\mathrm{tot,ST}}&=&\Phi_{\mathrm{strong,ST}}+\Phi_{\mathrm{weak,ST}}\nonumber\\
&=&\frac{2\pi n_e}{3}\sqrt{\frac{2}{\pi k_BT}}\left\{k_BT~\rho_{\mathrm{max}}^2\left(1-\exp\left[-\frac{v_{\mathrm{ST}}^2\left(\rho_{\mathrm{max}}\right)}{2k_BT}\right]\right)+E_1\left[\frac{v_{\mathrm{ST}}^2\left(\rho_{\mathrm{max}}\right)}{2k_BT}\right]\left(n_{\alpha}^2-n_{\beta}^2\right)^2\right\}.\nonumber\\
& &
\end{eqnarray}

\noindent Under the assumption 

\begin{equation}
\exp\left[-\frac{v_{\mathrm{ST}}^2\left(\rho_{\mathrm{max}}\right)}{2k_BT}\right]\approx 1-\frac{v_{\mathrm{ST}}^2\left(\rho_{\mathrm{max}}\right)}{2k_BT},
\end{equation}

\noindent the final Standard Theory form of the diagonal matrix element of the collision operator becomes therefore

\begin{equation}\label{fing}
\Phi_{\mathrm{tot,ST}}\approx \frac{2\pi n_e}{3}\sqrt{\frac{2}{\pi k_BT}}\left\{1+E_1\left[\frac{\left(n_{\alpha}^2-n_{\beta}^2\right)^2}{3k_BT\rho_{\mathrm{max}}^2}\right]\right\}\left(n_{\alpha}^2-n_{\beta}^2\right)^2,
\end{equation}

\noindent which is the expression of Griem \cite{GRIEM59}. It is also possible to integrate first on velocity and then on impact parameter (see Appendix \ref{appendixb}). 

Since most of the collisions are weak and correspond to $\rho\gg\rho_{w}$, where $\rho_w$ is the Weisskopf radius, they are therefore the main part of the broadening. The strong collisions correspond to $\rho<\rho_w$ and their contribution to the electron broadening represents usually less than 20 \% \cite{GRIEM59,GRIEM62}. 

\subsection{Case of penetrating collisions}\label{subsec42}

In low-density plasmas, the dominant contribution comes from long-range, distant collisions, for which the standard dipole approximation is not in question. For those close encounters which penetrate the wavefunction extent, the interaction is softened. Indeed, as mentioned above we have, for the emitter-perturber interaction energy \cite{ALEXIOU01}:

\begin{equation}
V(t)\approx\frac{1}{|\mathbf{R}(t)-\mathbf{r}(t)|}-\frac{1}{|\mathbf{R}(t)|},
\end{equation}

\noindent where $\mathbf{R}$ and $\mathbf{r}$ are the emitter and perturbing electron positions respectively and $V=0$ for $\mathbf{r}(t)=\mathbf{0}$, while this would diverge in the dipole approximation. Hence for close encounters, for which penetration occurs and the dipole approximation fails, we have a smaller (softer) interaction. It can happen, if the perturber velocity is high enough, that this softening changes the collision from a strong one to a weak one. This is particularly the case with almost head-on collisions, where a divergent interaction in Standard Theory actually gives a zero result when penetration is accounted for. At larger impact parameters, Standard Theory and Penetration Standard Theory give the same (small) value. At very small $\rho$, the latter gives 0, while Standard Theory diverges. These differences persist until about the relevant wavefunction extent. Except for the very small impact parameter regime, these differences are important if Penetration Standard Theory stays perturbative. Otherwise, the approximation $\langle\langle\alpha\beta|\{1-S_aS_b^{\dag}\}|\alpha'\beta'\rangle\rangle=1$ is thought to be an appropriate one for non perturbative behavior. This quantity oscillates around unity when unitarity breaks down. The relevance of penetration is then seen most clearly if the shielding length becomes small, so that a sizeable part of the impact parameter phase space is within the wavefunction extent. What happens in the extreme limit where the shielding length becomes less than the wavefunction extent, \emph{i.e.} for high densities or high principal quantum numbers is difficult to answer in detail. However, we may expect to find the usual Stark trends reversed and large deviations from Standard Theory.

Our new expression of the collision operator including penetration (see Eq.(\ref{newfor}) is easy to compute and facilitates the study and the accounting for penetrating collisions. It is interesting to see that the function $g$ behaves like $\ln\left(\rho_{\mathrm{max}}\right)$ (as in the standard theory without penetration effects) for high-enough values of the upper cutoff $\rho_{\mathrm{max}}$. Since the penetration standard theory is convergent for impact parameters as low as zero, there is no need for a minimum cutoff $\rho_{\mathrm{min}}$ (even though cutoffs on $v$ and $\rho$ should be introduced normally to avoid a violation of the perturbation theory). 

The determination of $v_{\mathrm{p}}(\rho)$, as in the Standard Theory the determination of $\rho_{\mathrm{p}}^2(v)$, amounts to solving:

\begin{eqnarray}
v_{\mathrm{p}}^2(\rho)&=&\frac{2}{3\rho^2}\left\{\delta_{\ell_{\beta},\ell_{\beta'}}\delta_{m_{\beta},m_{\beta'}}\sum_{\alpha''}\mathbf{r}_{\alpha\alpha''}\mathbf{r}_{\alpha''\alpha'}\left(1-\Delta\left(\frac{2\rho}{n_{\alpha}};n_{\alpha},\ell_{\alpha},\ell_{\alpha''}\right)\right)\left(1-\Delta\left(\frac{2\rho}{n_{\alpha}};n_{\alpha},\ell_{\alpha''},\ell_{\alpha'}\right)\right)\right.\nonumber\\
& &+\delta_{\ell_{\alpha},\ell_{\alpha'}}\delta_{m_{\alpha},m_{\alpha'}}\sum_{\beta''}\mathbf{r}_{\beta\beta''}\mathbf{r}_{\beta''\beta'}\left(1-\Delta\left(\frac{2\rho}{n_{\beta}};n_{\beta},\ell_{\beta''},\ell_{\beta'}\right)\right)\left(1-\Delta\left(\frac{2\rho}{n_{\beta}};n_{\beta},\ell_{\beta'},\ell_{\beta''}\right)\right)\nonumber\\
& &\left.-2\mathbf{r}_{\alpha\alpha'}\mathbf{r}_{\beta'\beta}\left(1-\Delta\left(\frac{2\rho}{n_{\alpha}};n_{\alpha},\ell_{\alpha},\ell_{\alpha'}\right)\right)\left(1-\Delta\left(\frac{2\rho}{n_{\beta}};n_{\beta},\ell_{\beta'},\ell_{\beta}\right)\right)\right\}.
\end{eqnarray}

\noindent Although this may be done separately for each matrix element, yielding therefore a specific $v_{\mathrm{min}}$ for each matrix element, we wish to keep the discussion on the same level as Standard Theory, which does not employ matrix-element dependent cutoffs. The quantity $v^2$ can be simplified as

\begin{eqnarray}\label{app1}
v_{\mathrm{p},\mathrm{app}}^{2}\left(\rho\right)&=&\frac{2}{3\rho^2}\left\{\delta_{\ell_{\beta},\ell_{\beta'}}\delta_{m_{\beta},m_{\beta'}}\sum_{\alpha''}\mathbf{r}_{\alpha\alpha''}\mathbf{r}_{\alpha''\alpha'}\left(1-\exp\left[-\frac{\left(2\rho /n_{\alpha}\right)^2}{2\chi_{n_{\alpha},\ell_{\alpha},\ell_{\alpha''}}^2}\right]\right)\left(1-\exp\left[-\frac{\left(2\rho /n_{\alpha}\right)^2}{2\chi_{n_{\alpha},\ell_{\alpha''},\ell_{\alpha'}}^2}\right]\right)\right.\nonumber\\
& &+\delta_{\ell_{\alpha},\ell_{\alpha'}}\delta_{m_{\alpha},m_{\alpha'}}\sum_{\beta''}\mathbf{r}_{\beta\beta''}\mathbf{r}_{\beta''\beta'}\left(1-\exp\left[-\frac{\left(2\rho /n_{\beta}\right)^2}{2\chi_{n_{\beta},\ell_{\beta'},\ell_{\beta''}}^2}\right]\right)\left(1-\exp\left[-\frac{\left(2\rho /n_{\beta}\right)^2}{2\chi_{n_{\beta},\ell_{\beta''},\ell_{\beta'}}^2}\right]\right)\nonumber\\
& &\left.-2\mathbf{r}_{\alpha\alpha'}\mathbf{r}_{\beta'\beta}\left(1-\exp\left[-\frac{\left(2\rho /n_{\alpha}\right)^2}{2\chi_{n_{\alpha},\ell_{\alpha},\ell_{\alpha'}}^2}\right]\right)\left(1-\exp\left[-\frac{\left(2\rho /n_{\beta}\right)^2}{2\chi_{n_{\beta},\ell_{\beta'},\ell_{\beta}}^2}\right]\right)\right\}.
\end{eqnarray}

\noindent where $\chi_{n,\ell,\ell'}$ is given by Eq. (\ref{chinl}). We can also replace the function $\chi_{n,\ell,\ell'}$ by its average over $\ell$ \cite{PAIN19}:

\begin{equation}
\bar{\chi}_n=\sum_{\ell=0}^{n-1}\chi_{n,\ell,\ell'}=\sqrt{\frac{\pi}{2}}~\frac{28n^2+n+6}{24n},
\end{equation}

\noindent which gives

\begin{eqnarray}\label{app2}
v_{\mathrm{p},\mathrm{app,2}}^{2}\left(\rho\right)&=&\frac{2}{3\rho^2}\left\{\delta_{\ell_{\beta},\ell_{\beta'}}\delta_{m_{\beta},m_{\beta'}}\sum_{\alpha''}\mathbf{r}_{\alpha\alpha''}\mathbf{r}_{\alpha''\alpha'}\left(1-\exp\left[-\frac{\left(2\rho /n_{\alpha}\right)^2}{2\bar{\chi}_{n_{\alpha}}^2}\right]\right)^2\right.\nonumber\\
& &+\delta_{\ell_{\alpha},\ell_{\alpha'}}\delta_{m_{\alpha},m_{\alpha'}}\sum_{\beta''}\mathbf{r}_{\beta\beta''}\mathbf{r}_{\beta''\beta'}\left(1-\exp\left[-\frac{\left(2\rho /n_{\beta}\right)^2}{2\bar{\chi}_{n_{\beta}}^2}\right]\right)^2\nonumber\\
& &\left.-2\mathbf{r}_{\alpha\alpha'}\mathbf{r}_{\beta'\beta}\left(1-\exp\left[-\frac{\left(2\rho /n_{\alpha}\right)^2}{2\bar{\chi}_{n_{\alpha}}^2}\right]\right)\left(1-\exp\left[-\frac{\left(2\rho /n_{\beta}\right)^2}{2\bar{\chi}_{n_{\beta}}^2}\right]\right)\right\}.
\end{eqnarray}

\noindent Using the sum rule (see Ref. \cite{VARSHA}, Eq. (3) p. 153):

\begin{equation}
\sum_{\psi,\kappa}(-1)^{p-\psi+q-\kappa}\ThreeJ{a}{p}{q}{-\alpha}{\psi}{\kappa}\ThreeJ{p}{q}{a'}{-\psi}{-\kappa}{\alpha'}=\frac{(-1)^{a+\alpha}}{(2a+1)}\delta_{a,a'}\delta_{\alpha,\alpha'},
\end{equation}

\noindent one gets

\begin{equation}
\sum_{j}\mathbf{r}_{ij}\mathbf{r}_{jk}=\frac{9}{4}n_i^2\left(n_i^2-\ell_i^2-\ell_i-1\right)\delta_{i,k},
\end{equation}

\noindent and Eq. (\ref{app1}) can be put in the form

\begin{eqnarray}\label{app2bis}
v_{\mathrm{p},\mathrm{app,2}}^{2}\left(\rho\right)&=&\frac{2}{3\rho^2}\left[\delta_{\alpha,\alpha'}\delta_{\beta,\beta'}\frac{9}{4}n_{\alpha}^2\left(n_{\alpha}^2-\ell_{\alpha}^2-\ell_{\alpha}-1\right)\left(1-\exp\left[-\frac{\left(2\rho /n_{\alpha}\right)^2}{2\bar{\chi}_{n_{\alpha}}^2}\right]\right)^2\right.\nonumber\\
& &+\delta_{\alpha,\alpha'}\delta_{\beta,\beta'}\frac{9}{4}n_{\beta}^2\left(n_{\beta}^2-\ell_{\beta}^2-\ell_{\beta}-1\right)\left(1-\exp\left[-\frac{\left(2\rho /n_{\beta}\right)^2}{2\bar{\chi}_{n_{\beta}}^2}\right]\right)^2\nonumber\\
& &\left.-2\mathbf{r}_{\alpha\alpha'}\mathbf{r}_{\beta'\beta}\left(1-\exp\left[-\frac{\left(2\rho /n_{\alpha}\right)^2}{2\bar{\chi}_{n_{\alpha}}^2}\right]\right)\left(1-\exp\left[-\frac{\left(2\rho /n_{\beta}\right)^2}{2\bar{\chi}_{n_{\beta}}^2}\right]\right)\right].
\end{eqnarray}

\noindent We want to compare

\begin{equation}
\Phi_{\mathrm{strong},1}=n_e\int_0^{\rho_{\mathrm{max}}}2\pi\rho d\rho \int_0^{v_{\mathrm{p}}(\rho)}vf(v)\langle\langle\alpha\beta|\left\{1-S_{a}S_{b}^{\dag}\right\}(\rho,v)|\alpha'\beta'\rangle\rangle dv
\end{equation}

\noindent with

\begin{equation}
\langle\langle\alpha\beta|\left\{1-S_{a}S_{b}^{\dag}\right\}(\rho,v)|\alpha'\beta'\rangle\rangle=\frac{v_{\mathrm{p}}^2(\rho)}{v^2}
\end{equation}

\noindent and

\begin{equation}
\Phi_{\mathrm{strong},2}=n_e\int_0^{\rho_{\mathrm{max}}}2\pi\rho d\rho\int_0^{v_{\mathrm{p}}(\rho)}vf(v)dv.
\end{equation}

\noindent The quantity $\Phi_{\mathrm{strong},1}$ represents the contribution of the penetrating collision operator in the regime of strong collisions. The latter is convergent in that regime, but this does not mean that the results are correct. On the other hand, $\Phi_{\mathrm{strong},2}$ represents the way the strong collisions should be treated. If $\Phi_{\mathrm{strong},1}$ and $\Phi_{\mathrm{strong},2}$ differ significantly in that region, this means that the Penetration Standard Theory is not applicable in the strong-collision regime, although it is convergent, and that $\Phi_{\mathrm{strong},1}$ must be replaced by $\Phi_{\mathrm{strong},2}$. We get

\begin{equation}\label{phimatstrong1}
\Phi_{\mathrm{strong},1}=2\pi n_e\sqrt{\frac{2}{\pi k_BT}}\int_0^{\rho_{\mathrm{max}}}\rho~ v_{\mathrm{p}}(\rho)^2\left\{1-\exp\left[-\frac{v_{\mathrm{p}}^2(\rho)}{2k_BT}\right]\right\}d\rho.
\end{equation}

\noindent and

\begin{equation}\label{phimatstrong2}
\Phi_{\mathrm{strong},2}=\frac{4\pi n_e}{3}\sqrt{\frac{2}{\pi k_BT}}\left\{-\frac{3}{2}\int_0^{\rho_{\mathrm{max}}}\rho\exp\left[-\frac{v_{\mathrm{p}}^2(\rho)}{2k_BT}\right]\left(v_{\mathrm{p}}(\rho)^2+2k_BT\right)d\rho+\frac{3}{2}k_BT\rho_{\mathrm{max}}^2\right\}
\end{equation}

\noindent The weak-collision part is

\begin{equation}\label{phimatweak}
\Phi_{\mathrm{weak}}=2\pi n_e\sqrt{\frac{2}{\pi k_BT}}\int_0^{\rho_{\mathrm{max}}}\rho~v_{\mathrm{p}}^2(\rho)\exp\left[-\frac{v_{\mathrm{p}}^2(\rho)}{2k_BT}\right]d\rho.
\end{equation}

\noindent We can also compare with the strong-collision contribution in the framework of the Standard Theory:

\begin{equation}\label{phimatst}
\Phi_{\mathrm{strong,ST}}=n_e\int_0^{\rho_{\mathrm{max}}}2\pi\rho d\rho \int_0^{v_{\mathrm{ST}}(\rho)}vf(v)dv
\end{equation}

\noindent with

\begin{eqnarray}
v_{\mathrm{ST}}^{2}\left(\rho\right)=\frac{2}{3\rho^2}\left[\sum_{\alpha''}\mathbf{r}_{\alpha\alpha''}\mathbf{r}_{\alpha''\alpha'}+\sum_{\beta''}\mathbf{r}_{\beta\beta''}\mathbf{r}_{\beta''\beta'}-2\mathbf{r}_{\alpha\alpha'}\mathbf{r}_{\beta'\beta}\right],
\end{eqnarray}

\noindent which can be put in the form

\begin{eqnarray}
v_{\mathrm{ST}}^{2}\left(\rho\right)=\frac{2}{3\rho^2}\left[\frac{9}{4}n_{\alpha}^2\left(n_{\alpha}^2-\ell_{\alpha}^2-\ell_{\alpha}-1\right)+\frac{9}{4}n_{\beta}^2\left(n_{\beta}^2-\ell_{\beta}^2-\ell_{\beta}-1\right)-2\mathbf{r}_{\alpha\alpha'}\mathbf{r}_{\beta'\beta}\right].
\end{eqnarray}

\noindent Our results can be checked using sum rules (see Appendix \ref{appendixc}). As discussed in Sec. \ref{subsec41}, integrating on velocities first, and then on impact parameters gives the same result, but the domain has to be split into

\begin{equation}
\int_0^{v_{\mathrm{ST}}\left(\rho_{\mathrm{max}}\right)}vf(v)dv\int_0^{\rho_{\mathrm{max}}}2\pi\rho d\rho+\int_{v_{\mathrm{ST}}\left(\rho_{\mathrm{max}}\right)}^{\infty}vf(v)dv\int_0^{\rho_{\mathrm{ST}}(v)}2\pi\rho d\rho
\end{equation}

\noindent with

\begin{eqnarray}
\rho_{\mathrm{ST}}^2\left(v\right)&=&\frac{2}{3v^2}\left[\sum_{\alpha''}\mathbf{r}_{\alpha\alpha''}\mathbf{r}_{\alpha''\alpha'}+\sum_{\beta''}\mathbf{r}_{\beta\beta''}\mathbf{r}_{\beta''\beta'}-2\mathbf{r}_{\alpha\alpha'}\mathbf{r}_{\beta'\beta}\right]\nonumber\\
&=&\frac{2}{3v^2}\left[\frac{9}{4}n_{\alpha}^2\left(n_{\alpha}^2-\ell_{\alpha}^2-\ell_{\alpha}-1\right)+\frac{9}{4}n_{\beta}^2\left(n_{\beta}^2-\ell_{\beta}^2-\ell_{\beta}-1\right)-2\mathbf{r}_{\alpha\alpha'}\mathbf{r}_{\beta'\beta}\right].
\end{eqnarray}

\noindent The quantity $\Phi_{\mathrm{strong,ST}}$ is equal to

\begin{equation}
\Phi_{\mathrm{strong,ST}}=2n_e\sqrt{2\pi k_BT}~\rho_{\mathrm{max}}^2\left\{1-\exp\left[-\frac{v_{\mathrm{ST}}^2\left(\rho_{\mathrm{max}}\right)}{2k_BT}\right]\right\}
\end{equation}

\noindent and represents the strong collisions term in the Standard Theory (see section \ref{subsec41}).

As pointed out by Alexiou \cite{ALEXIOU05} in the context of the interpretation of the H$_{\alpha}$ experiments mentioned above \cite{BODDEKER93,BUSCHER02}, Penetration Standard Theory yields larger widths for the weak-collision contribution, but a much smaller strong-collision contribution. This increases the relative proportion of the phase space that is reliably computed (weak collisions) compared to those approximated (strong collisions). 

\begin{figure}
\begin{center}
\includegraphics[width=13cm]{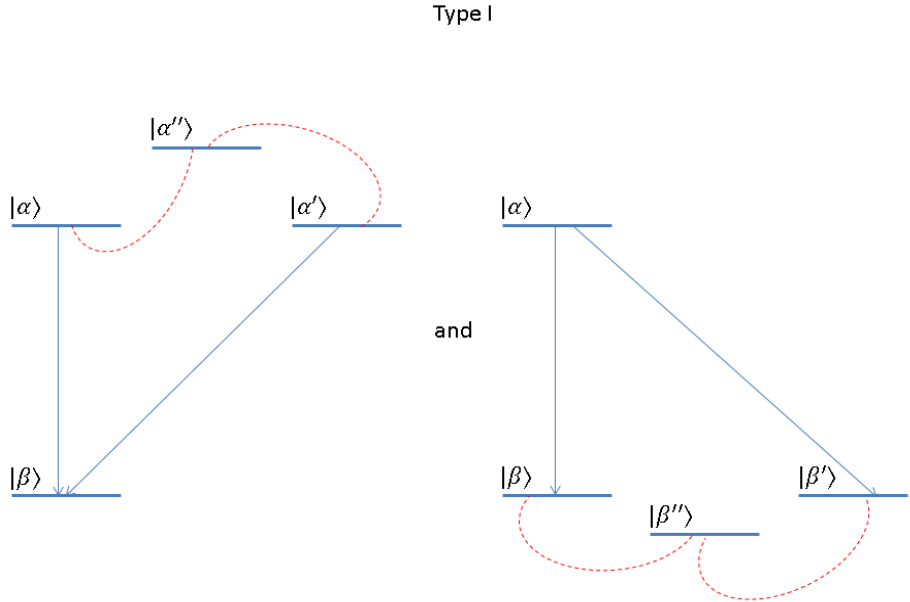}
\end{center}
\caption{Atomic schemes corresponding to nondiagonal terms of the collision operator (type I).}\label{fig1}
\end{figure}

\begin{figure}
\begin{center}
\includegraphics[width=13cm]{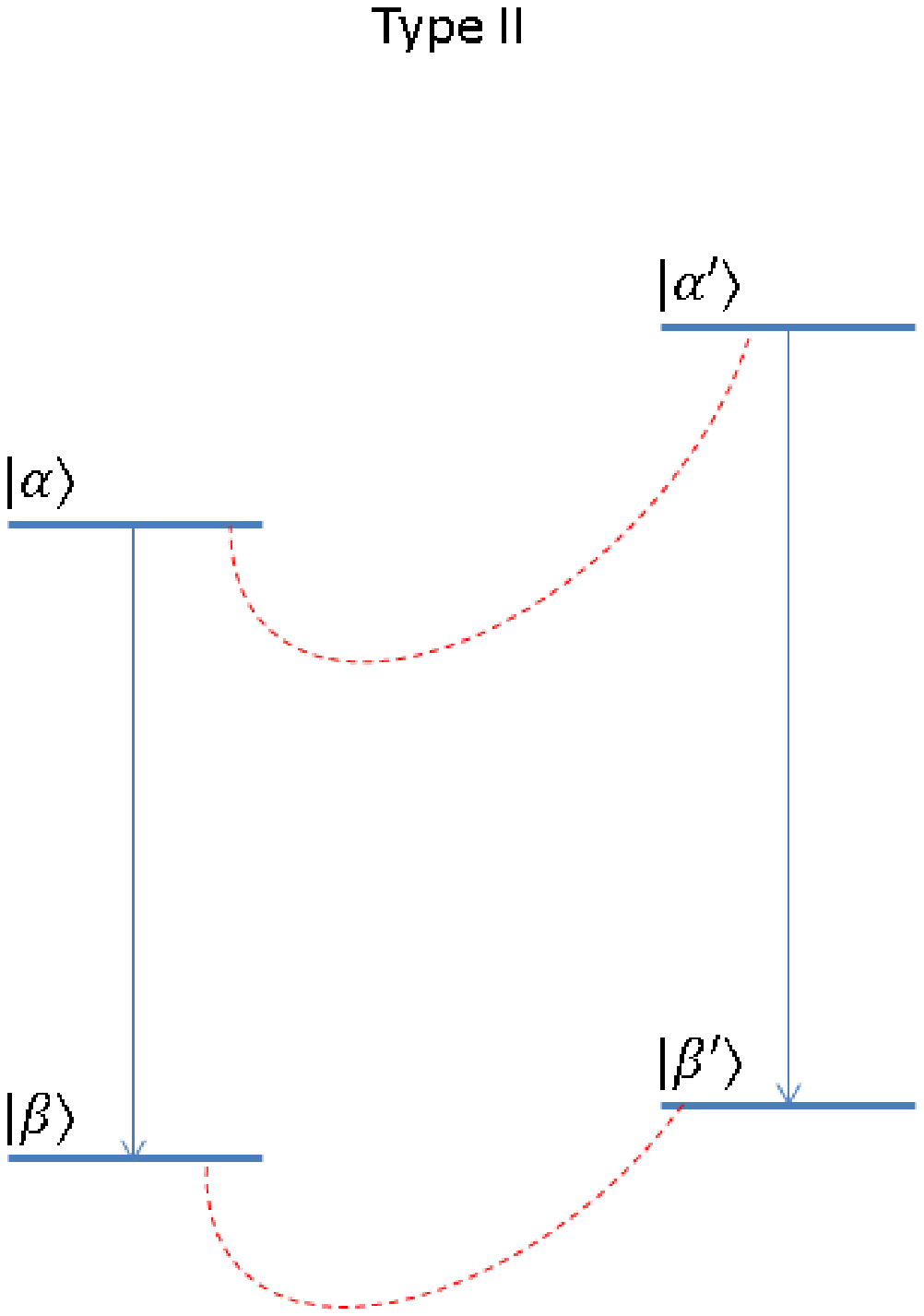}
\end{center}
\caption{Atomic schemes corresponding to interference terms of the collision operator (type II).}\label{fig2}
\end{figure}

\begin{figure}
\begin{center}
\includegraphics[width=13cm]{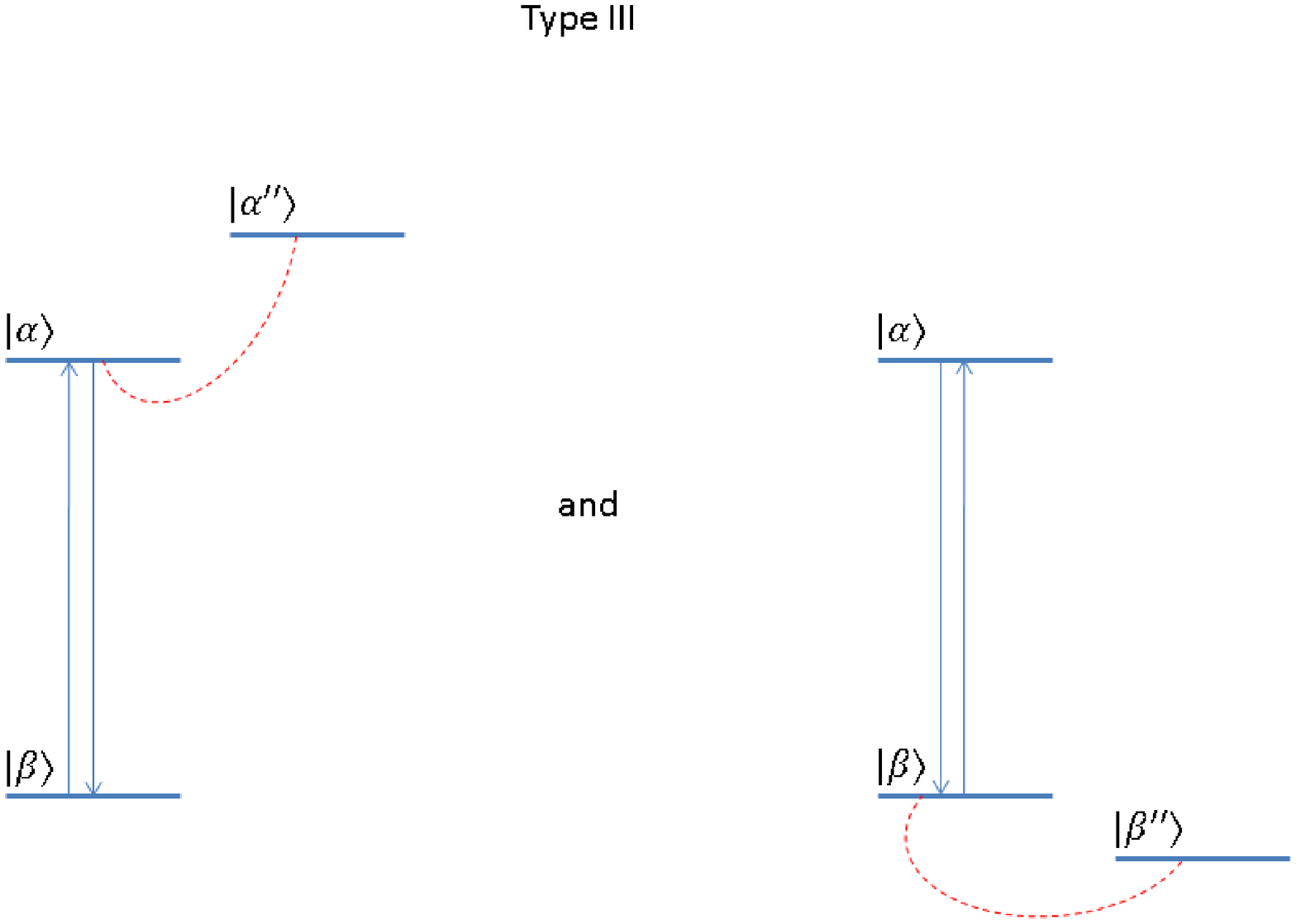}
\end{center}
\caption{Atomic schemes corresponding to diagonal terms of the collision operator (type III).}\label{fig3}
\end{figure}

\begin{figure}
\begin{center}
\includegraphics[width=10cm]{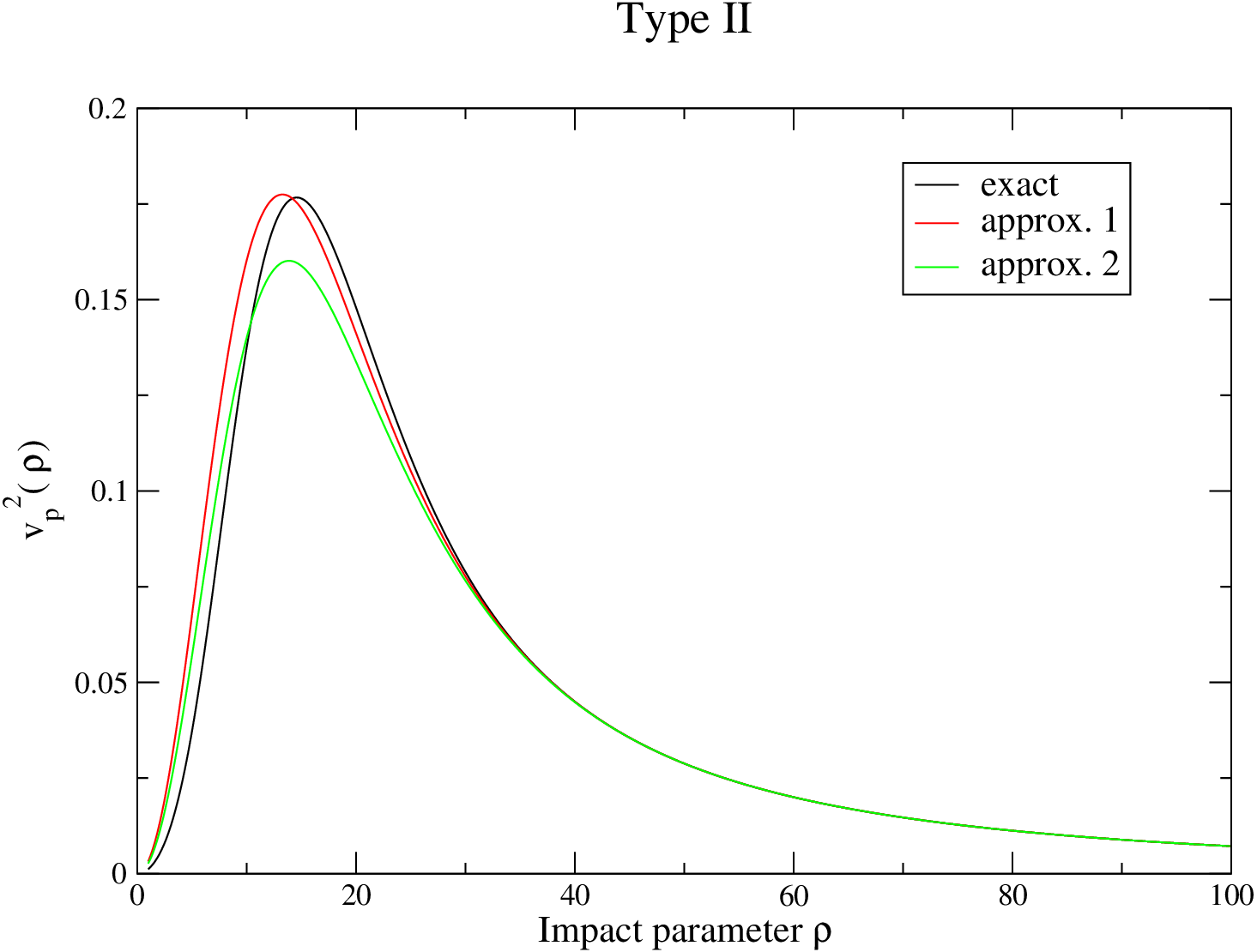}
\end{center}
\caption{Comparison between the exact computation of $v_{\mathrm{p}}^2(\rho)$ and the approximate expressions $v_{\mathrm{p},\mathrm{app1}}$ (see Eq. (\ref{app1})) and $v_{\mathrm{p},\mathrm{app},2}$ (see Eq. (\ref{app2bis})) for $n_{\alpha}=4$, $\ell_{\alpha}=3$, $m_{\alpha}=2$, $\ell_{\alpha'}=2$ and $m_{\alpha'}=1$, $n_{\beta}=3$, $\ell_{\beta}=2$, $m_{\beta}=2$, $\ell_{\beta'}=1$ and $m_{\beta'}=1$ (Type II).}\label{fig4}
\end{figure}

\begin{figure}
\begin{center}
\includegraphics[width=10cm]{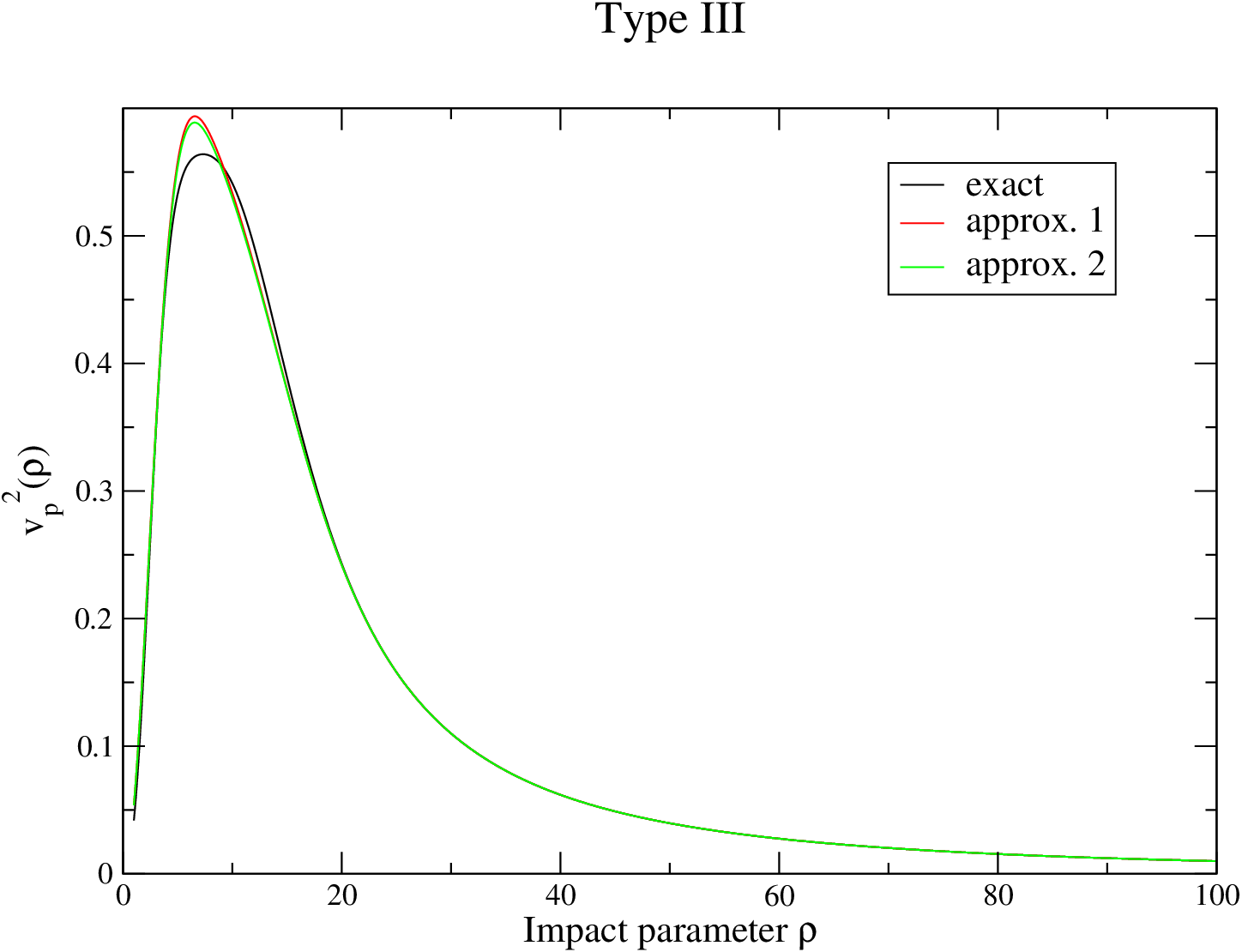}
\end{center}
\caption{Comparison between the exact computation of $v_{\mathrm{p}}^2(\rho)$ and the approximate expressions $v_{\mathrm{p},\mathrm{app}}$ (see Eq. (\ref{app1})) and $v_{\mathrm{p},\mathrm{app},2}$ (see Eq. (\ref{app2bis})) for $n_{\alpha}=3$, $\ell_{\alpha}=1$, $m_{\alpha}=1$, $\ell_{\alpha'}=1$ and $m_{\alpha'}=1$, $n_{\beta}=2$, $\ell_{\beta}=0$, $m_{\beta}=0$, $\ell_{\beta'}=0$ and $m_{\beta'}=0$ (Type III).}\label{fig5}
\end{figure}

\begin{figure}
\begin{center}
\includegraphics[width=10cm]{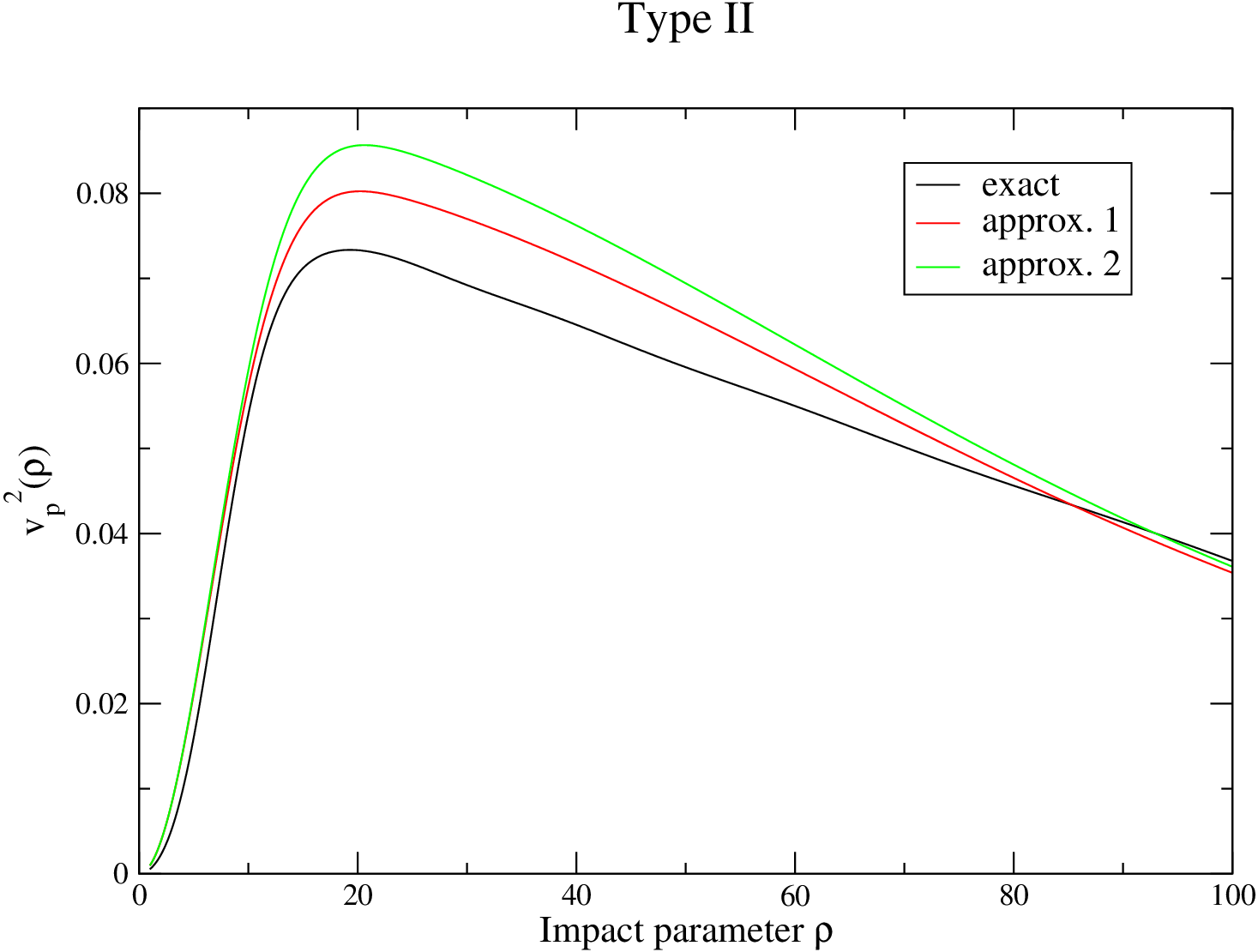}
\end{center}
\caption{Comparison between the exact computation of $v_{\mathrm{p}}^2(\rho)$ and the approximate expressions $v_{\mathrm{p},\mathrm{app}}$ (see Eq. (\ref{app1})) and $v_{\mathrm{p},\mathrm{app},2}$ (see Eq. (\ref{app2bis}) for $n_{\alpha}=8$, $\ell_{\alpha}=3$, $m_{\alpha}=2$, $\ell_{\alpha'}=2$ and $m_{\alpha'}=1$, $n_{\beta}=3$, $\ell_{\beta}=2$, $m_{\beta}=2$, $\ell_{\beta'}=1$ and $m_{\beta'}=1$ (Type II).}\label{fig6}
\end{figure}

\begin{figure}
\begin{center}
\includegraphics[width=10cm]{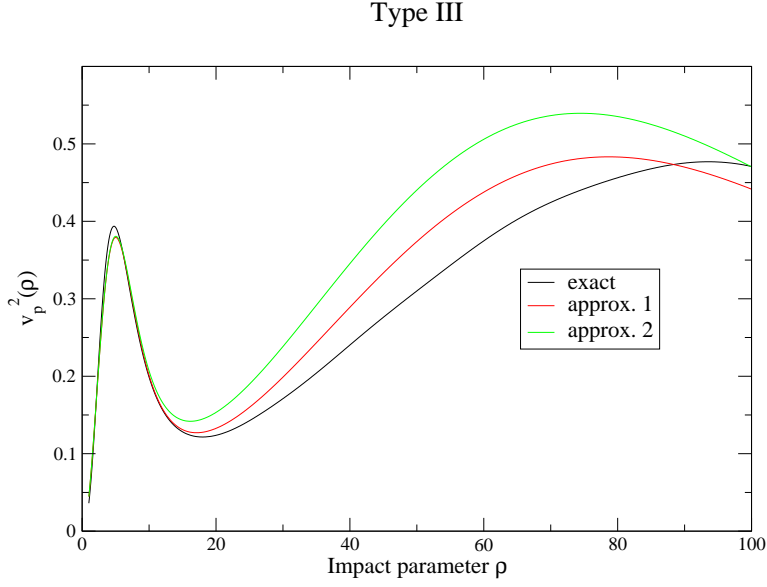}
\end{center}
\caption{Comparison between the exact computation of $v_{\mathrm{p}}^2(\rho)$ and the approximate expressions $v_{\mathrm{p},\mathrm{app}}$ (see Eq. (\ref{app1})) and $v_{\mathrm{p},\mathrm{app},2}$ (see Eq. (\ref{app2bis})) for $n_{\alpha}=8$, $\ell_{\alpha}=1$, $m_{\alpha}=1$, $\ell_{\alpha'}=1$ and $m_{\alpha'}=1$, $n_{\beta}=2$, $\ell_{\beta}=0$, $m_{\beta}=0$, $\ell_{\beta'}=0$ and $m_{\beta'}=0$ (Type III).}\label{fig7}
\end{figure}

\begin{figure}
\begin{center}
\includegraphics[width=10cm]{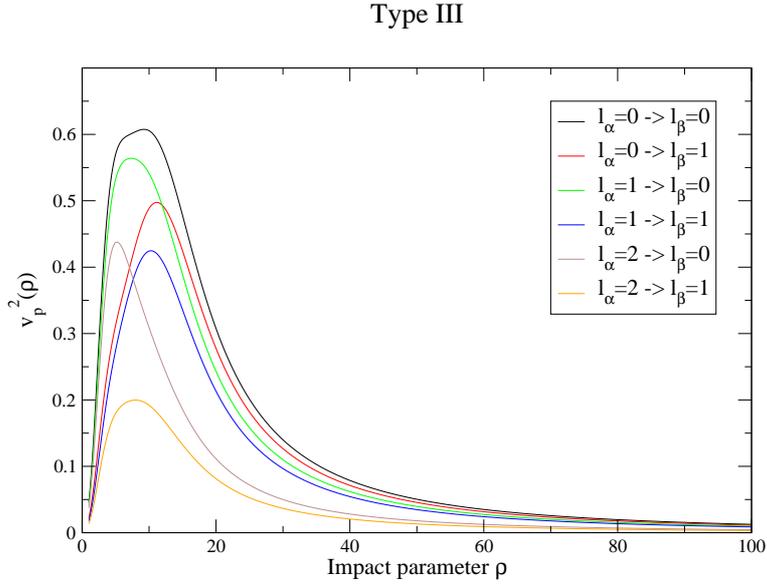}
\end{center}
\caption{Exact computation of $v_{\mathrm{p}}^2(\rho)$ for all channels included in $n_{\alpha}=3\rightarrow n_{\beta}=2$ (Type III).}\label{fig8}
\end{figure}

\begin{figure}
\begin{center}
\includegraphics[width=10cm]{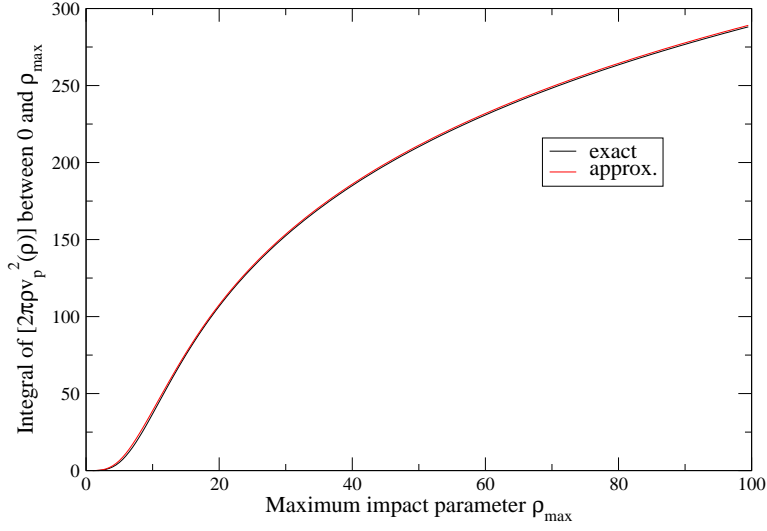}
\end{center}
\caption{Comparison between exact expression of $\int_0^{\rho_{\mathrm{max}}}2\pi\rho v_{\mathrm{p}}^2(\rho)d\rho$ and the approximate expression $v_{\mathrm{p},\mathrm{app}}$ (see Eq. (\ref{app1})) in the framework of Penetration Theory for $n_{\alpha}=3$, $\ell_{\alpha}=2$, $m_{\alpha}=1$, $n_{\beta}=2$, $\ell_{\beta}=1$, $m_{\beta}=0$, $\ell_{\alpha'}=m_{\alpha'}=1$ and $\ell_{\beta'}=m_{\beta'}=0$ as a function of $\rho_{\mathrm{max}}$.}\label{fig9}
\end{figure}

\begin{figure}
\begin{center}
\includegraphics[width=10.cm]{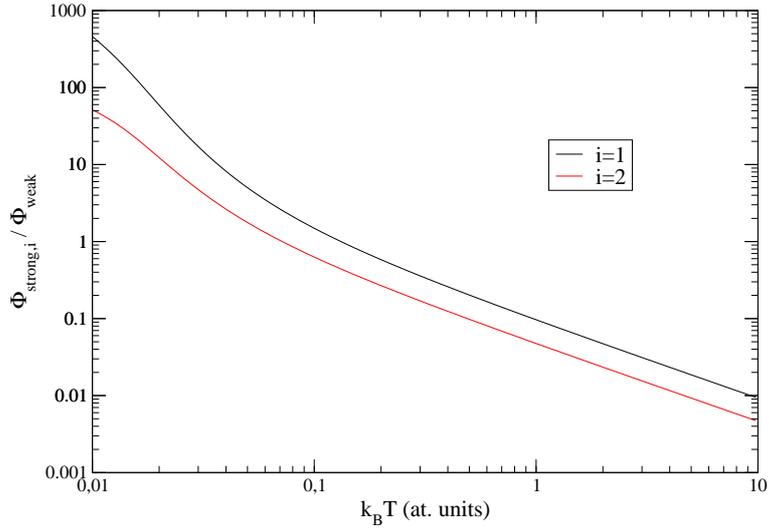}
\end{center}
\caption{Ratio $\Phi_{\mathrm{strong},1}/\Phi_{\mathrm{weak}}$ (see Eqs. (\ref{phimatstrong1}) and (\ref{phimatweak})) and $\Phi_{\mathrm{strong},2}/\Phi_{\mathrm{weak}}$ (see Eqs. (\ref{phimatstrong2}) and (\ref{phimatweak})) as functions of temperature (atomic units) for $n_{\alpha}$=3, $\ell_{\alpha}$=2, $n_{\beta}$=2, $\ell_{\beta}$=1 and $\rho_{\mathrm{max}}=n_{\alpha}^2$.}\label{fig10}
\end{figure}

\begin{figure}
\begin{center}
\includegraphics[width=10.cm]{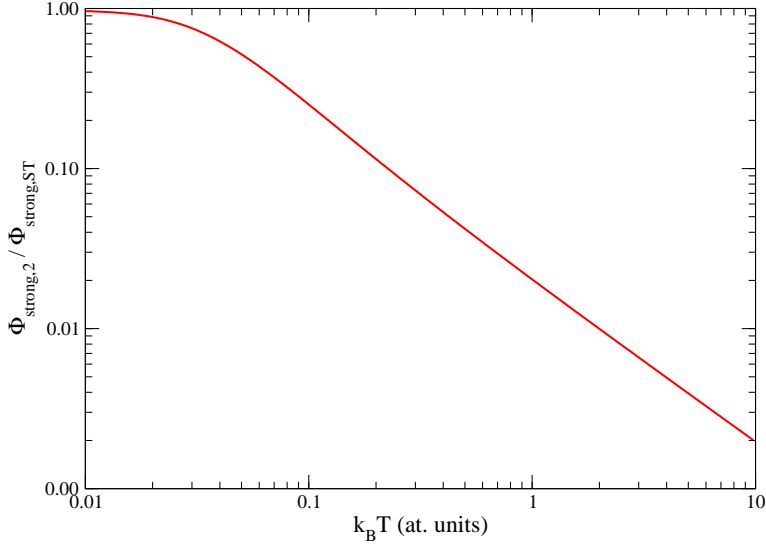}
\end{center}
\caption{Ratio $\Phi_{\mathrm{strong},2}/\Phi_{\mathrm{strong,ST}}$ (see Eq. (\ref{phimatst})) as a function of temperature (atomic units) for $n_{\alpha}$=3, $\ell_{\alpha}$=2, $n_{\beta}$=2, $\ell_{\beta}$=1 and $\rho_{\mathrm{max}}=n_{\alpha}^2$.}\label{fig11}
\end{figure}

Figures \ref{fig1}, \ref{fig2} and \ref{fig3} represent the different schemes (transitions or channels) for the modelling of the collision operator. The first one (type I, see Fig. \ref{fig1}) corresponds to nondiagonal terms of the operator and does not exist in the case of hydrogen; the second one (type II, see Fig. \ref{fig2}) to interference terms and the third one (type III, see Fig. \ref{fig3}) to diagonal terms. The first one gives a zero contribution in the case of the hydrogen atom. Figures \ref{fig3}, \ref{fig4}, \ref{fig5} and \ref{fig6} display comparisons between the exact computation of $v_{\mathrm{p}}^2(\rho)$ and the approximate expressions $v_{\mathrm{p},\mathrm{app1}}$ (see Eq. (\ref{app1}) and $v_{\mathrm{p},\mathrm{app},2}$ (see Eq. (\ref{app2bis})) for different channels (type II for figures \ref{fig4} and \ref{fig6}) and type III for figures \ref{fig5} and \ref{fig7}. For both types, two values of the principal quantum number were chosen: $n_{\alpha}=3$ (Figs. \ref{fig4} and \ref{fig5}) and $n_{\alpha}=8$ (Figs. \ref{fig6} and \ref{fig7}). We can see that the approximate form (\ref{app1}) is very close to the exact results, and that the cruder approximation (\ref{app2bis}) still has a rather satisfactory accuracy. In the case of a high value of the principal quantum number, the agreement is not as good, especially for impact parameters larger than 20. Figure \ref{fig8} displays the exact computation of $v_{\mathrm{p}}^2(\rho)$ for all channels included in $n_{\alpha}=3\rightarrow n_{\beta}=2$ (Type III). We can see that the dispersion is very important; therefore, it would probably not be relevant to calculate and average collision operator between the two shells $n_{\alpha}$ and $n_{\beta}$. A comparison between exact expression of $\int_0^{\rho_{\mathrm{max}}}2\pi\rho v_{\mathrm{p}}^2(\rho)d\rho$ and the approximate expression $v_{\mathrm{p},\mathrm{app}}$ (see Eq. (\ref{app1}) in the framework of Penetration Theory for $n_{\alpha}=3$, $\ell_{\alpha}=2$, $n_{\beta}=2$, $\ell_{\beta}=1$ as a function of $\rho_{\mathrm{max}}$ is presented in figure \ref{fig9}. Here also, the agreement is excellent. Figure \ref{fig10} shows the ratios $\Phi_{\mathrm{strong},1}/\Phi_{\mathrm{weak}}$ (see Eqs. (\ref{phimatstrong1}) and (\ref{phimatweak})) and $\Phi_{\mathrm{strong},2}/\Phi_{\mathrm{weak}}$ (see Eqs. (\ref{phimatstrong1}) and (\ref{phimatweak})) as functions of temperature (atomic units) for $\rho_{\mathrm{max}}=n_{\alpha}^2$. Since $\Phi_{\mathrm{strong},1}$ and $\Phi_{\mathrm{strong},2}$ differ significantly in that region (especially for very low temperatures), the Penetration Standard Theory must not be applied in the regime of strong collisions, although it is convergent, and  $\Phi_{\mathrm{strong},1}$ must be replaced by $\Phi_{\mathrm{strong},2}$. Note that the strong collisions become comparable (and even larger) to weak collisions for temperatures smaller than $\approx 1$ eV. The variation of the ratio $\Phi_{\mathrm{strong},2}/\Phi_{\mathrm{strong,ST}}$ (see Eq. (\ref{phimatst})) as a function of temperature for the same conditions (excitation channel, maximum impact parameter, \emph{etc.} as Fig. \ref{fig10}) is represented in figure \ref{fig11} and reveals that the strong collisions are probably largely overestimated in the Standard Theory. Figure \ref{fig11} displays a simplified schematic representation of three different collisional regimes: screened, weak and strong in the ($\rho$,$v$) plane, $\rho$ being the impact parameter and $v$ the electron velocity.

When the temperature is high, unitarity-violating collisions are not significant and the Standard Theory ``strong collision term'' is misrepresented, as it arises from very weak, penetrating collisions. The qualitative behavior of Penetration Standard Theory is to be expected: at very low densities, the phase space inside the wave-function extent is unable to compete with the large impact parameter phase space, hence the relative importance must tend to 0 as the density decreases. Similarly, for very high densities, all of the phase space tends to be completely inside the wavefunction extent, and this means a decreasing relative strong contribution, as even slow collisions are softened more and more by penetration. Hence a maximum is expected for Penetration Standard Theory. Generally, the weak collision contribution to Penetration Standard Theory is larger than the corresponding Standard Theory contribution, because of the larger Penetration Standard Theory weak collision phase space, while the strong collision contribution to Penetration Standard Theory is much smaller than the corresponding strong collision contribution to Standard Theory. This is why the relative strong collision width is smaller in the Penetration Standard Theory, which in turn means increased confidence in the final result. 

In this work, we have considered that perturbing electrons pass the radiating atom as free particles. In reality, they move in the dipole potential (hydrogen atom possesses permanent electric dipole moment). Oks proposed to overcome that assumption \cite{OKS15}. He also proposed a more accurate definition of the so-called Weisskopf radius, different from of one used by Griem. This might be important because the choices of the Weisskopf radius and the strong collision constant are interrelated. Oks found that the latter refinements increase the electron broadening, especially for warm dense plasmas.

\section{Conclusion}\label{sec5}

A semi-classical model for the electron broadening operator including the effect of penetrating collisions on isolated lines of hydrogen, \emph{i.e.} collisions in which the incoming electron enters the extent of bound-electron wavefunctions, was developed by Alexiou and Poqu\'erusse. The corresponding formalism is rather complex and involves recursive calculations and Bessel and Bickley-Naylor functions. We derived an approximate expression for the collision operator, which is very simple, easy to compute and accurate. Such a formula should also help to improve the understanding of strong collisions and the limits of standard theory. However, one has to keep in mind the fact that, in the Penetration Standard Theory, the collision operator is convergent whatever the value of the maximum impact parameter, even when penetration theory is not valid anymore. Therefore, we discussed the problem of strong collisions when penetration effects are taken into account and found that applying the penetration theory even for very low values of the impact parameter (\emph{i.e.} when the density is low and/or the temperature is high) may lead to overestimate the contribution of strong collisions to the line broadening.

\appendix
\section{A. The interaction picture: from the general case to hydrogen in the Standard Theory}\label{appendixa}

\subsection{Collision operator in the interaction picture for neutral emitter}

\noindent The electron collision operator can be developed with respect to the interaction $V(t)$ in a perturbation series:

\begin{eqnarray}\label{dys}
\Phi_{ab}&=&n_e\int_0^{\infty}vf(v)dv\int_0^{\infty}2\pi\rho d\rho\left\{\frac{1}{\hbar^2}\left[-\int_{-\infty}^{\infty}\tilde{V}_a(t)dt\int_{-\infty}^{\infty}\tilde{V}_b(t)dt\right.\right.\nonumber\\
& &+\int_{-\infty}^{\infty}\tilde{V}_a(t)dt\int_{-\infty}^{t}\tilde{V}_a(t')dt'\left.\left.+\int_{-\infty}^{\infty}\tilde{V}_b(t)dt\int_{-\infty}^{t}\tilde{V}_b(t')dt'\right]\right\},
\end{eqnarray}

\noindent where, in the interaction picture:

\begin{equation}
\tilde{V}(t)=\exp\left[i\frac{\hat{H}t}{\hbar}\right]~V(t)~\exp\left[-i\frac{\hat{H}t}{\hbar}\right],
\end{equation}

\noindent $\hat{H}$ being the Hamiltonian. Under the straight-path assumption (valid only for a neutral emitter), the perturbation $V(t)$ produced by the collision with an electron has the form

\begin{equation}
V(t)=\frac{\mathbf{r}.(\boldsymbol{\rho}+\mathbf{v}t)}{\left(\rho^2+v^2t^2\right)^{3/2}},
\end{equation}

\noindent where $\mathbf{r}$ is the radius vector of an atomic electron. In Eq. (\ref{dys}), the first order does not contribute, because the average of $\left\{V\right\}$ over all directions of the vectors $\boldsymbol{\rho}$ and $\mathbf{v}$ is zero. For the second order, we have \cite{SOBELMAN72a,SOBELMAN72b}:

\begin{equation}
\left\{V(t)V(t')\right\}=\frac{\mathbf{r}.\mathbf{r}}{3}\frac{(\rho^2+v^2tt')}{\left(\rho^2+v^2t^2\right)^{3/2}\left(\rho^2+v^2t'^2\right)^{3/2}}.
\end{equation}

\noindent The second and third terms of expression (\ref{dys}) can be calculated using:

\begin{eqnarray}\label{seco}
& &\langle\langle\alpha\beta|\int_{-\infty}^{\infty}V_a(t)dt\int_{-\infty}^{t}V_a(t')dt'|\alpha'\beta'\rangle\rangle\nonumber\\
&=&\int_{-\infty}^{\infty}dt\int_{-\infty}^{\infty}dt'\exp\left[i\left(\epsilon_{\alpha\alpha''}t+\epsilon_{\alpha''\alpha'}t'\right)\right]\sum_{\alpha''}\langle\alpha|V_a(t)|\alpha''\rangle\langle\alpha''|V_b(t)|\alpha'\rangle\delta_{\beta,\beta'}\nonumber\\
&=&\frac{1}{3}\sum_{\alpha''}\mathbf{r}_{\alpha\alpha''}\mathbf{r}_{\alpha''\alpha'}\int_{-\infty}^{\infty}dt\int_{-\infty}^{\infty}dt'\frac{(\rho^2+v^2tt')}{\left(\rho^2+v^2t^2\right)^{3/2}\left(\rho^2+v^2t'^2\right)^{3/2}}\exp\left[i\left(\epsilon_{\alpha\alpha''}t+\epsilon_{\alpha''\alpha'}t'\right)\right],
\end{eqnarray}

\noindent where $\epsilon_{ij}=\epsilon_j-\epsilon_j$ represents the difference between the energies of states $i$ and $j$. By introducing the dimensionless variables

\begin{equation}
\left\{
\begin{array}{l}
z_1=\frac{\rho}{v}\epsilon_{\alpha\alpha''}\\
z_2=\frac{\rho}{v}\epsilon_{\alpha'\alpha''}\\
x_1=\frac{vt}{\rho}\\
x_2=\frac{vt'}{\rho},
\end{array}.
\right.
\end{equation}

\noindent equation (\ref{seco}) becomes

\begin{equation}
\frac{1}{3}\sum_{\alpha''}\mathbf{r}_{\alpha\alpha''}\mathbf{r}_{\alpha''\alpha'}\frac{1}{\rho^2v^2}~J(z_1,z_2),
\end{equation}

\noindent where

\begin{equation}
J(z_1,z_2)=\int_{-\infty}^{\infty}dx_1\int_{-\infty}^{\infty}dx_2\frac{(1+x_1x_2)}{\left(1+x_1^2\right)^{3/2}\left(1+x_2^2\right)^{3/2}}\exp\left[i\left(z_1x_1-z_2x_2\right)\right]=A(z_1,z_2)+iB(z_1,z_2).
\end{equation}

\noindent The summation over $\alpha''$ is restricted to the states of the level $a$ and neglecting the perturbation due to all the other levels. The values $\mathbf{r}_{\alpha\alpha''}$ are not zero only for neighboring Stark components $\alpha, \alpha''$. At $\alpha=\alpha'$, $z_1=z_2=z$ and at $\alpha\ne\alpha'$, $z_1=-z_2=z$. Let us denote the corresponding integrals $A_+(z)$, $B_+(z)$ and $A_-(z)$, $B_-(z)$. The real part $A_{\pm}(z)$ can be expressed in terms of modified Bessel functions as

\begin{equation}
A_{\pm}(z)=z^2\left[K_1^2(z)\pm K_0^2(z)\right],
\end{equation}

\noindent where $K_0$ and $K_1$ are Bessel functions of the second kind \cite{ABRAMOWITZ64} (sometimes called Basset functions or Macdonald functions). One must, in general, compute $B_{\pm}(z)$ from a dispersion relation, making use of the fact that $A$ and $B$ are real and imaginary parts of the same complex function (where $\mathcal{P}$ indicates Cauchy principal value \cite{GRIEM74,COOPER69}):

\begin{equation}
B_{\pm}(z)=\frac{2|z|}{\pi}\mathcal{P}\int_{-\infty}^{\infty}\frac{A_{\pm}(z')}{z^2-z'^2}dz'.
\end{equation}

\noindent Their asymptotic behavior for large $z\gg 1$ yields

\begin{equation}
\left\{
\begin{array}{l}
A_{\pm}(z)\approx\pi|z|\exp\left[-2|z|\right],\\
B_{\pm}(z)\approx\pi/4z,
\end{array}
\right.
\end{equation}

\noindent and for small $z\ll 1$:

\begin{equation}
\left\{\begin{array}{l}
A_{\pm}(z)\approx 1,\\
B_{\pm}(z)\approx 0.
\end{array}\right.
\end{equation}

\subsection{Case of hydrogen}

For hydrogen, the exponential functions disappear in Eq. (\ref{seco}) because $\epsilon_{\alpha\alpha''}=0$ and $\epsilon_{\alpha\alpha''}=0$. One has therefore (as in the previous $z\ll 1$ case): $A_{\pm}(z)=1$ and $B_{\pm}(z)=0$.

\appendix
\section{B. Strong collisions in the Standard Theory: integrating first on velocity and then on impact parameter}\label{appendixb}

It is worth mentioning that it is possible to interchange the integrations, \emph{i.e.} to integrate first on velocity $v$, and then on impact parameter $\rho$. For the strong-collision part, this means

\begin{equation}
\Phi_{\mathrm{strong,ST}}=n_e\int_0^{\rho_{\mathrm{max}}}2\pi\rho' d\rho'\int_0^{v_{\mathrm{ST}}(\rho')}vf(v)dv,
\end{equation}

\noindent where $v_{\mathrm{ST}}(\rho)$ is given by Eq. (\ref{vst}). The integration over velocities gives

\begin{equation}
\int_0^{v_{\mathrm{ST}}(\rho)}vf(v)dv=\frac{1}{(k_BT)^{3/2}}\sqrt{\frac{2}{\pi}}\left\{2(k_BT)^2-\frac{2k_BT}{3\rho^2}\exp\left[-\frac{v_{\mathrm{ST}}^2\left(\rho\right)}{2k_BT}\right]\times\left[\left(n_{\alpha}^2-n_{\beta}^2\right)^2+3k_BT\rho^2\right]\right\}
\end{equation} 

\noindent and thus

\begin{equation}
\Phi_{\mathrm{strong,ST}}=2n_e\sqrt{2\pi k_BT}~\rho_{\mathrm{max}}^2\left\{1-\exp\left[-\frac{v_{\mathrm{ST}}^2\left(\rho_{\mathrm{max}}\right)}{2k_BT}\right]\right\}.
\end{equation}

\noindent As concerns the weak-collision part

\begin{equation}
\Phi_{\mathrm{weak,ST}}=n_e\int_0^{\rho_{\mathrm{max}}}2\pi\rho'd\rho'\int_{v_{\mathrm{ST}}(\rho')}^{\infty}vf(v)dv,
\end{equation}

\noindent we have

\begin{equation}
\int_{v_{\mathrm{ST}}(\rho)}^{\infty}vf(v)\frac{2}{3v^2\rho^2}\left(n_{\alpha}^2-n_{\beta}^2\right)^2dv=\frac{2}{3\rho^2}\sqrt{\frac{2}{\pi k_BT}}\exp\left[-\frac{v_{\mathrm{ST}}^2\left(\rho\right)}{2k_BT}\right]\left(n_{\alpha}^2-n_{\beta}^2\right)^2
\end{equation}

\noindent and thus

\begin{equation}
\Phi_{\mathrm{weak,ST}}=\frac{2\pi n_e}{3}\sqrt{\frac{2}{\pi k_BT} }E_1\left[\frac{v_{\mathrm{ST}}^2\left(\rho\right)}{2k_BT}\right]\left(n_{\alpha}^2-n_{\beta}^2\right)^2,
\end{equation}

\noindent which yields

\begin{eqnarray}
\Phi_{\mathrm{tot,ST}}&=&\Phi_{\mathrm{strong,ST}}+\Phi_{\mathrm{weak,ST}}\nonumber\\
&=&\frac{2\pi n_e}{3}\sqrt{\frac{2}{\pi k_BT}}\left\{k_BT\rho_{\mathrm{max}}^2\left(1-\exp\left[-\frac{v_{\mathrm{ST}}^2\left(\rho_{\mathrm{max}}\right)}{2k_BT}\right]\right)+E_1\left[\frac{v_{\mathrm{ST}}^2\left(\rho_{\mathrm{max}}\right)}{2k_BT}\right]\left(n_{\alpha}^2-n_{\beta}^2\right)^2\right\}\nonumber\\
& &
\end{eqnarray}

\noindent which is exactly Eq. (\ref{phitotfirst}). Making the substitution

\begin{equation}
\exp\left[-\frac{v_{\mathrm{ST}}^2\left(\rho_{\mathrm{max}}\right)}{2k_BT}\right]\approx 1-\frac{v_{\mathrm{ST}}^2\left(\rho_{\mathrm{max}}\right)}{2k_BT},
\end{equation}

\noindent we get

\begin{equation}
\Phi_{\mathrm{strong,ST}}\approx\frac{2\pi n_e}{3}\sqrt{\frac{2}{k_BT}}\left(n_{\alpha}^2-n_{\beta}^2\right)^2
\end{equation}

\noindent and

\begin{equation}
\Phi_{\mathrm{tot,ST}}\approx \frac{2n_e}{3}\sqrt{\frac{2\pi }{k_BT}}\left\{1+E_1\left[\frac{v_{\mathrm{ST}}^2\left(\rho_{\mathrm{max}}\right)}{2k_BT}\right]\right\}\left(n_{\alpha}^2-n_{\beta}^2\right)^2,
\end{equation}

\noindent which is exactly Eq. (\ref{fing}). 

\section{C. checking the matrix elements using sum rules}\label{appendixc}

In order to check the calculations, it is useful to resort to sum rules.

\noindent $\bullet$ For instance, a simplification of the expression giving the total strength between shells $n$ and $n'$ has been obtained by McLean \cite{MCLEAN34, WATSON06B, HEY09} using recursion relations between Gauss hypergeometric functions:

\begin{align}\label{C}
S_{nn'}&=\sum_{\ell=0}^{n-1}\sum_{\ell'=0}^{n'-1}\left[\left(\Rnl{n'\ell+1}{n\ell}\right)^2\delta_{\ell',\ell+1}+\left(\Rnl{n'\ell-1}{n\ell}\right)^2\delta_{\ell',\ell-1}\right]\\
&=\frac{(2nn')^6(n'-n)^{2n+2n'-5}}{Z^2(n+n')^ {2n+2n'+4}}\left\{\left[\phantom{x\!\!}_2 F_1(-n'+1,-n;1;X)\right)^2-\left(\phantom{x\!\!}_2 F_1(-n',-n+1;1;X)\right]^2\right\},
\end{align}

\noindent where $X=-4nn'/\left(n-n'\right)^2$. One should have

\begin{equation}
\sum_{\alpha,\beta}\mathbf{r}_{\alpha\beta}\mathbf{r}_{\alpha\beta}=\sum_{\ell_{\alpha}=0}^{n_{\alpha}-1}\sum_{m_{\alpha}=-\ell_{\alpha}}^{\ell_{\alpha}}\sum_{\ell_{\beta}=0}^{n_{\beta}-1}\sum_{m_{\beta}=-\ell_{\beta}}^{\ell_{\beta}}\mathbf{r}_{\alpha\beta}\mathbf{r}_{\alpha\beta}=S_{n_{\alpha}n_{\beta}},
\end{equation}

\noindent where $\mathbf{r}_{\alpha\beta}$ is given by Eq. (\ref{rij}).

\noindent $\bullet$ Another interesting check is provided by the average collision operator \cite{CASINI95,GILLERON19}

\begin{align}
C_{nn'}&=\frac{9}{4}\left[\left(n^2-n'^2\right)^2-n^2-n'^2\right].
\end{align}

\noindent Defining

\begin{equation}
\Upsilon_{\alpha,\alpha',\beta,\beta'}=\delta_{\beta\beta'}\sum_{\alpha''}\mathbf{r}_{\alpha\alpha''}.\mathbf{r}_{\alpha''\alpha'}+\delta_{\alpha\alpha'}\sum_{\beta''}\mathbf{r}_{\beta'\beta''}.\mathbf{r}_{\beta''\beta}-2\mathbf{r}_{\alpha\alpha'}.\mathbf{r}_{\beta'\beta},
\end{equation}

\noindent one has

\begin{equation}
\sum_{\alpha,\alpha',\beta,\beta'}\Upsilon_{\alpha,\alpha',\beta,\beta'}\times\mathbf{r}_{\beta\alpha}\mathbf{r}_{\alpha'\beta'}=S_{n_{\alpha}n_{\beta}}\times C_{n_{\alpha}n_{\beta}}.
\end{equation}

\clearpage

{\bf acknowledgments:} This work is dedicated to the memory of Hans Griem.

\end{document}